\documentclass[reprint,aip,jcp,graphicx,twocolumn]{revtex4-1}

\usepackage{bm}
\usepackage{dcolumn}
\usepackage{graphicx}
\usepackage{amsmath}

\newcommand{\fig}[1]{Figure~\ref{#1}}
\newcommand{\tab}[1]{Table~\ref{#1}}
\newcommand{\eq}[1]{Equation~\ref{#1}}

\newcolumntype{d}[1]{D{.}{.}{#1}}

\draft 

\begin{document}


\title{Implicit Solvation Using the Superposition Approximation (IS-SPA): Extension to Polar Solutes in Chloroform.} 



\author{Peter T. Lake}
\altaffiliation{These authors contributed equally to this work.}
\affiliation{Department of Chemistry, Oklahoma State University}

\author{Max A. Mattson}
\altaffiliation{These authors contributed equally to this work.}
\affiliation{Department of Chemistry, Colorado State University}

\author{Martin McCullagh}
\email[]{martin.mccullagh@okstate.edu}
\affiliation{Department of Chemistry, Oklahoma State University}


\date{\today}

\begin{abstract}
Efficient, accurate, and adaptable implicit solvent models remain a significant challenge in the field of molecular simulation.  A recent implicit solvent model, IS-SPA, based on approximating the mean solvent force using the superposition approximation, provides a platform to achieve these goals.  IS-SPA was originally developed to handle non-polar solutes in the TIP3P water model but can be extended to accurately treat polar solutes in other polar solvents.  In this manuscript, we demonstrate how to adapt IS-SPA to include the treatment of solvent orientation and long ranged electrostatics in a solvent of chloroform.  The orientation of chloroform is approximated as that of an ideal dipole aligned in a mean electrostatic field.  The solvent--solute force is then considered as an averaged radially symmetric Lennard-Jones component and a multipole expansion of the electrostatic component through the octupole term.  Parameters for the model include atom-based solvent density and mean electric field functions that are fit from explicit solvent simulations of independent atoms or molecules.  Using these parameters, IS-SPA accounts for asymmetry of charge solvation and reproduces the explicit solvent potential of mean force of dimerization of two oppositely charged Lennard-Jones spheres with high fidelity.  Additionally, the model more accurately captures the effect of explicit solvent on the monomer and dimer configurations of alanine dipeptide in chloroform than a generalized Born or constant density dielectric model.  The current version of the algorithm is expected to outperform explicit solvent simulations for aggregation of small peptides at concentrations below 150 mM, well above the typical experimental concentrations for these materials.
\end{abstract}

\pacs{}

\maketitle 

\section{Introduction}
\label{intro}

The solvation of polar molecules changes the conformations they sample and the higher order structures they form relative to that found in vacuum.  Self-assembling peptides, for example, achieve different macroscopic structure and properties under different solvent environments.\cite{Shen:1995} Computational tools used to predict this behavior are limited due to (1) the length and time scales achievable using all-atom simulations and (2) the the lack of solvent transferability in coarse-grained models.  A computationally cheap, thermodynamically accurate, and solvent adaptable implicit solvent model will have an immediate impact on studying these processes.

Molecular simulations play an important role at discerning the mechanism of aggregation and self-assembly of peptide-based materials.  Experimental approaches often lack the multiscale resolution necessary to determine macroscopic structure as well as the underlying molecular driving forces.  All-atom molecular dynamics (aaMD) simulations have been employed to help elucidate the latter component.\cite{McCullagh:2008,Lee:2011,Anderson:2018,Weber:2020}  aaMD, however, cannot readily sample the large length- and time-scales necessary to capture macroscopic self-assembly.  Additionally, these methods struggle with the low concentration of solutes typically used in self-assembly experiments.\cite{Yan:2010}  Top-down coarse-grained (CG) simulations and mesoscale models have been used to investigate peptide aggregation but their tie to atomistic detail is often obscured.\cite{Mansbach:2017,Frederix:2011,Frederix:2015,Tuttle:2015}  Bottom-up CG models provide a rigorous tie between scales but often lack transferability.  Recent efforts have attempted to alleviated the transferability issue but have not been widely adopted in molecular simulations.\cite{Sanyal:2018,Jin:2019} Recently, a multiscale model for peptide assembly has been proposed that uses a combination of top-down CG models and aaMD simulations but a rigorous connections between scales is not guaranteed.\cite{Zhao:2019} Thus, it remains an open challenge in the field of molecular simulation to achieve computationally efficient and thermodynamically consistent bridge between all-atom and CG models.\cite{Colombo:2007}


Implicit solvent models provide a necessary bridge between aaMD simulations and CG models and an appealing platform for self-assembly prediction.  Previous simulation studies of peptide aggregation using implicit solvation are limited.\cite{Shell:2008,Ishizuka:2010} The limited application of these models is due to the inaccuracies of the computationally feasible models\cite{Shell:2008} and the significant computational expense of more accurate models.\cite{Ishizuka:2010}  Recent developments have attempted to bridge the gap between the computationally expensive and cheap models but, as of yet, the Goldilocks implicit solvent model does not exist.\cite{Remsing:2011,Remsing:2016,Gao:2020} Additionally, much of the development of implicit solvent models focus on water as the solvent and the portability to other solvents is not considered.  

Our recently developed implicit solvent model, implicit solvation using the superposition approximation (IS-SPA), provides a platform to achieve an accurate and computationally feasible implicit solvent model.\cite{Lake:2017}  This model builds on previous approaches that utilize the Kirkwood superposition approximation (SPA)\cite{Kirkwood:1935} to estimate the mean solvent force for a given solute configuration.\cite{Pellegrini:1995,Pellegrini:1996}  Initially developed for non-polar solutes in water, nothing about the underlying theory is specific to water as a solvent.  Indeed, the SPA is predicted be more accurate for a less polar solvent, such as chloroform.  

In this paper, we extend the development of IS-SPA to accurately and efficiently capture peptide monomer configurations and dimerization behavior in chloroform.  In the subsequent theory section, we set forth the underlying theory of IS-SPA and how it is adapted for (1) non-spherically symmetric solvent forces and (2) polar solutes in a polar solvent.  The accuracy and limitations of the approach are demonstrated on the solvation and dimerization of spherical charges in chloroform followed by the modeling of the monomeric conformations and dimerization behavior of alanine dipeptide (ADP) in chloroform.


\section{Theory}
\label{theory}

IS-SPA relies on estimating the average solvent force on atom $i$ given the position $\bm{R}^N$ of the $N$ solute atoms, which is given exactly when the solvent is spherically symmetric as
\begin{equation}
  \left\langle \bm{f}_i \right\rangle_\text{solv} = \rho \int \bm{f}_{i,\text{solv}}(\bm{r} - \bm{R}_i)\; g(\bm{r};\bm{R}^N) \;\text{d}\bm{r},
\label{exact}
\end{equation}
where $\rho$ is the solvent density, $g$ is the solvent distribution function, and $\bm{f}_{i,\text{solv}}$ is the solvent force.  The first generation of IS-SPA introduced two main approximations to estimate \eq{exact} on the fly: (1) the many-body solvent distribution function, $g(\bm{r};\bm{R}^N)$, is approximated using SPA with atomic radial distribution functions fit from the results of a single molecular configuration and (2) Monte Carlo (MC) integration is used to approximate the integral.\cite{Lake:2017}  This second point yields a simulation with the correct force on average, with the thermostat being responsible for removing any excess heat produced by the inexact calculation. This approach to an implicit solvent model was shown to capture physics of solvent interactions for non-polar association in water better than other implicit solvent models such as SASA and RISM.\cite{Lake:2017}  Subsequent sections detail the extensions of IS-SPA to account for the non-spherically symmetric solvent potentials, the long range electrostatic forces, and how the associated parameters are calculated.



\subsection{Inclusion of Non-spherically Symmetric Solvent Potentials}

The previous version of IS-SPA only considered the spherically symmetric Lennard-Jones solvent forces from the TIP3P model of water.  In the case of polar solutes in a solvent of chloroform, both the Coulomb and Lennard-Jones solvent--solute forces also depend on the orientation of the solvent relative to the solute atoms.  The mean solvent force found in \eq{exact} is changed to be
\begin{equation}
  \left\langle \bm{f}_i \right\rangle_\text{solv} = \rho \iint \bm{f}_{i,\text{solv}}(\bm{r} - \bm{R}_i,\bm{\Omega})\; g(\bm{r},\bm{\Omega};\bm{R}^N)\; \text{d}\bm{\Omega}\; \text{d}\bm{r},
\label{exact2}
\end{equation}
where $\bm{\Omega}$ represents the internal coordinates of the solvent molecule. We use a series of approximations to analytically integrate over $\bm{\Omega}$ such that only the position $\bm{r}$ of the solvent needs to be sampled. 

The first approximation we introduce is to presume that the only important internal and orientational degree of freedom of a solvent molecule is the alignment of its dipole moment and that the molecule is axially symmetric.  This reduces the twelve degrees of freedom in $\bm{\Omega}$ for a chloroform molecule to two, signified by $\bm{\hat{p}}$.  Next, the SPA is modified to account for the orientation of the dipole moment.  Namely,
\begin{equation}
    g(\bm{r},\bm{\hat{p}};\bm{R}^N)\simeq \frac{\prod_{i=1}^N g_i(|\bm{r}-\bm{R}_i|)\; P_i(\bm{\hat{p}};|\bm{r}-\bm{R}_i|)}{\int \prod_{i=1}^N P_i(\bm{\hat{p}};|\bm{r}-\bm{R}_i|)\; \text{d}\bm{\hat{p}}},
\end{equation}
where $P_i(\bm{\hat{p}};\bm{|\bm{r}-\bm{R}_i}|)$ is the probability distribution of the dipole moment at a given distance from atom $i$.  

Next, we approximate the distribution of the dipole moment to be that of a thermally ideal dipole in a constant electric field.  Each atom is presumed to produce a radially symmetric mean field along its separation vector, $E_i(\bm{r}) \bm{\hat{r}}$.  Thus,
\begin{equation}
    P_i(\bm{\hat{p}};|\bm{r}-\bm{R}_i|) \simeq \exp\left[-\frac{p}{T}\bm{\hat{p}}\cdot\bm{\hat{r}}\, E_{i}(|\bm{r}-\bm{R}_i|)\right],
\end{equation}
where $p$ is the static dipole moment of a chloroform molecule and $T$ is the temperature in units of energy.  In this sense, the orientation is dependent on the superposition of electric fields generated by each atom.   The only parameters that are needed for the model to predict the solvent distribution function are the atomic radial distribution functions $g_i(r)$ and the magnitude of the effective electric field from each atom $E_i(r)$. The latter is measured in simulation through the average polarization and related to the effective electric field using the Langevin function. 

\begin{figure}
\includegraphics[width=0.2\textwidth]{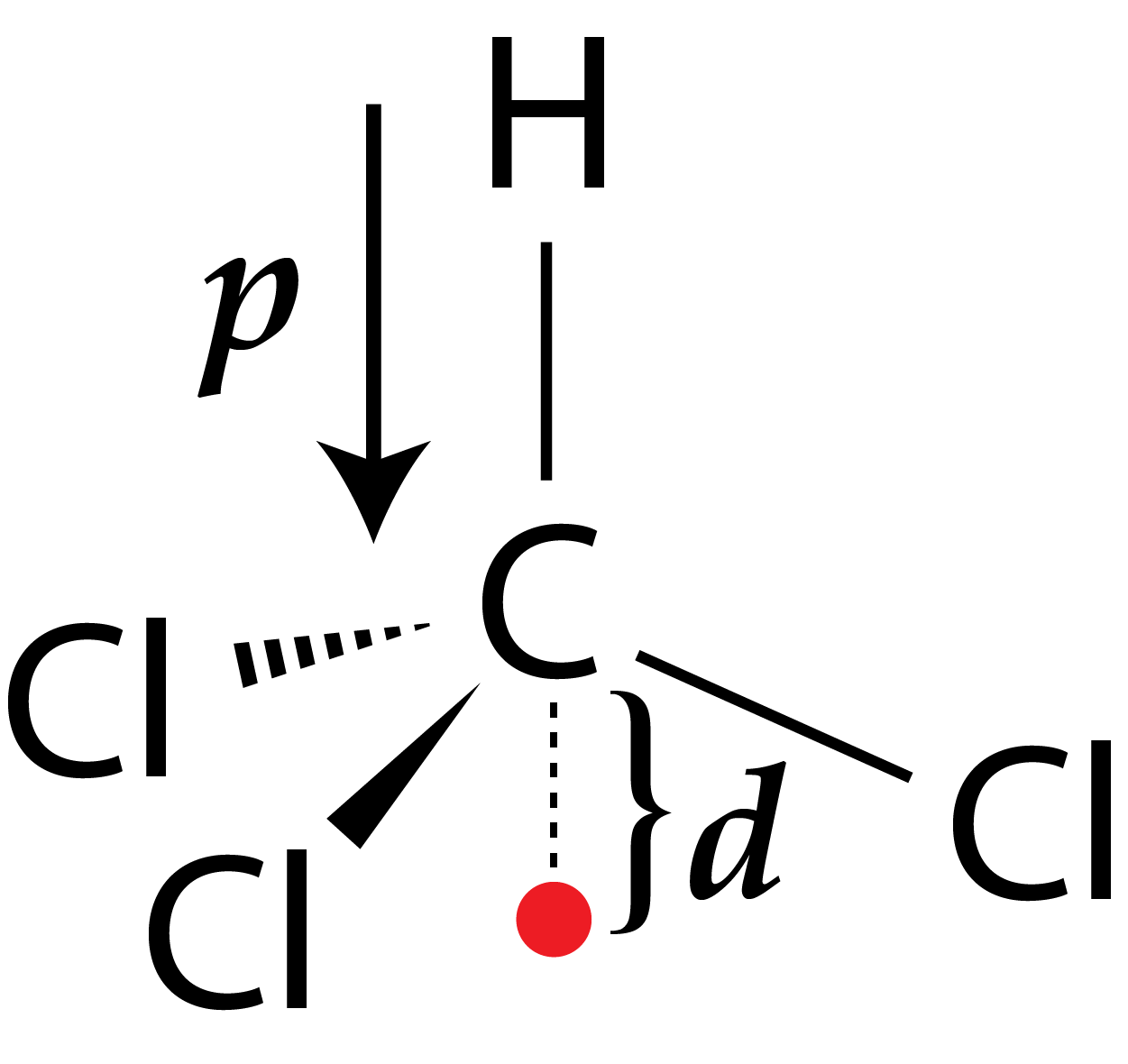}
\caption{\label{chloroform_schematic} Schematic of a chloroform molecule with the dipole moment labeled as $\bm{p}$ and the displacement of the center of force of the molecule labeled as $d$.}
\end{figure}
The solute--solvent force functions depend on all of the solvent degrees of freedom so must be simplified in a manner analogous to the distribution function.  We use the expansion of the axial multipole moments for the electrostatic force.  A description of the equations associated with the solvent force are found in the Supplementary Information.  For this, the multipole moments are calculated for the minimum energy structure of the solvent molecule in vacuum, assuming that the distribution of the rotation around the dipole moment is isotropic.  A free parameter, $d$, is introduced relating to the position of the molecular center relative to the carbon atom along the dipole of the molecule, with positive value being towards the chlorine atoms (\fig{chloroform_schematic}).  We choose a value of -0.21 \AA ~from the carbon atom in order to minimize the magnitude of the moments and to have the hexadecapole moment be zero, as shown in Figure SI2.     We then calculate the solvent electrostatic force to fourth order.

The simplification of the Lennard-Jones potential is not as immediately obvious.  It is possible to expand the potential analogous to the multipole expansion.  The problem with this approach is that the functions do not converge as quickly due to the rapidly varying potential energy as a function of orientation.  Instead, we consider the force to be radially symmetric, effectively truncating the expansion at zeroeth order.  We use the same definition of the solvent molecular center as with electrostatic force.  Instead of developing a function to describe the Lennard-Jones force, the average force that the solvent puts on each atom at a given distance along the separation vector is measured and histogrammed to be used in the simulations.


\subsection{Treatment of Long Ranged Electrostatics}

The inclusion of electrostatic forces requires handling their long range nature.  We calculate an analytic result for the long ranged interactions instead of proposing any approximations to simplify the calculation of the electrostatic solvent force.  We limit the discussion to non-periodic systems to avoid needing to use methods such as particle mesh Ewald.  This limitation still allows for calculating the monomer configurations and dimer PMFs discussed herein.

The approach to calculating the integral of the solvent force over all space is to divide the space into two parts.  One part is the union of spherical volumes within some cut off distance from the solute atoms, named the interaction volume.  The MC sampling is only performed within the interaction volume and the force from any MC point is calculated for all solute atoms.  Outside of the interaction volume, the Lennard-Jones force is set to zero, the solvent density is presumed to be that of the bulk fluid, and the polarization density is taken to be that found in a constant density dielectric (CDD).  A cut off distance of 12 \AA ~is chosen to define the interaction volume based on when the variations in density and polarization tend to these bulk values.

An analytic result for the force due to a CDD continuum is not immediately accessible for an arbitrary shape found for the interaction volume created by a molecule.  What is calculated readily is the result of the solvent force on atom $i$ from the solvent polarization produced by the second $j$ outside the interaction volume of the two atoms.  Namely,
\begin{widetext}
\begin{equation}
    \bm{f}^{\text{CDD}}_{ij}(r_{ij})= \begin{cases}
    0 & i=j\\
    -\frac{q_1q_2}{4\pi\epsilon_0} \left(1-\frac{1}{\epsilon} \right ) \frac{r_{ij}(8r_{\text{cut}}-3r_{ij})}{24r_{\text{cut}}^4}\bm{\hat{r}}_{ij} & i\ne j, r_{ij}<2r_{\text{cut}}\\
    -\frac{q_1q_2}{4\pi\epsilon_0} \left(1-\frac{1}{\epsilon} \right ) \frac{2}{3r_ij^2}\bm{\hat{r}}_{ij} & i\ne j, r_{ij}>2r_{\text{cut}}
  \end{cases}.
\label{analytic2}
\end{equation}
\end{widetext}
This pairwise force is added to the calculation of the direct solute--solute forces.  Since the law of superposition is an exact relation in the case of the CDD, adding all these pairwise interactions correctly calculates the force outside of the interaction volume, at the expense of also calculating the forces inside the interaction volume of the rest of the molecule but outside that of the given $ij$ pair.  This overcounted force is sampled along with the MC sampling within the molecular interaction volume and subtracted from the final forces.  Note that this force is non-zero even within the excluded volume of the atoms such that the MC integration needs to sample volumes that have zero density.

This exact approach to calculating the long range force introduces many inefficiencies to the model, including the need to sample within the excluded volume of the solute where the solvent forces are identically zero and that the force of each MC sampled point needs to be calculated for each solute atom regardless of separation distance.  These inefficiencies are a target for future approximations to simplify the calculations while introducing minimal error.


\subsection{Fit Parameters for Molecular Systems}
There are two parameter sets discussed in the above theory that need to be measured for the system: the atomic radial distribution functions and effective electric fields.  In the previous work on IS-SPA, it is shown that the SPA is not valid to describe the solvent distribution function of a molecule built upon the radial distribution functions of each separate atom.  Instead, it is better to fit the parameters to the solvent distribution function observed for the molecule.  The input for this method is achieved by running a simulation with the solute molecule pinned in space and sampling the solvent distribution.  The mean density and polarization of solvent is measured in a cubic grid with cells much smaller than the size of a solvent molecule, cubes of linear length 0.25 \AA ~in the case of chloroform.  Instead of restricting the fit to some functional form with a minimal number of free parameters, the functional form of the fit is a binned function of distance with bins of 0.1 \AA.

Since the underlying statistics for measuring the distribution function from a molecular dynamics simulation is counting statistics, the goodness of fit function that is minimized in the case of finding parameters for the atomic radial distribution function is related to the Poisson distribution.  Finding the most likely set of parameters to describe the explicit solvent simulation results leads to minimizing the function $\Upsilon_g$ with respect to each fit parameter,
\begin{equation}
    \Upsilon_g = \sum_{n\in \text{cells}} \left[\prod_{i=1}^N g_i(|\bm{r}_n-\bm{R}_i|) - g^{\text{obs}}_n \sum_{i=1}^N \ln g_i(|\bm{r}_n-\bm{R}_i|)\right],
    \label{fit_g}
\end{equation}
where the outer sum is over all cells in the measured distribution function, the inner sum and product are over all solute atoms, and $g^{\text{obs}}_n$ is the observed value of the distribution function of the cell. The Newton-Raphson method is applied to iteratively converge to the minimum in $\Upsilon_g$.

A different goodness of fit is needed for the case of finding the effective electric field generated by each atom in the solute.  The nature of the statistics associated with the distribution of polarizations of a solvent molecule is not as simple, but it can be approximated as being that of a dipole in a constant external field.  Finding the most likely set of parameters to describe the aaMD results leads to minimizing the function $\Upsilon_{EF}$ with respect to each fit parameter,
\begin{widetext}
\begin{equation}
  \Upsilon_{EF}=\sum_{n\in \text{cells}} N_n\left[ \ln \left( \frac{p}{T}|\bm{E}^{\text{mod}}_n| \right) -\ln \sinh \left( \frac{p}{T} |\bm{E}^{\text{mod}}_n| \right) + \frac{p}{T}\langle \bm{\hat{p}}\rangle_n\cdot \bm{E}^{\text{mod}}_n \right],
  \label{fit_p}
\end{equation}
\end{widetext}
where $N_n$ is the number of solvent molecules observed in the cell, $\langle \bm{\hat{p}}\rangle_n$ is the average unit vector of the dipole moment in the cell, and the model effective electric field $\bm{E}^{\text{mod}}_n$ is equal to
\begin{equation}
    \bm{E}^{\text{mod}}_n=\sum_{i=1}^N \frac{E_i(|\bm{r}_n-\bm{R}_i|)}{|\bm{r}_n-\bm{R}_i|}(\bm{r}_n-\bm{R}_i).
\end{equation}
In this sense, the polarization is the observable in the simulation and the effective electric field per atom is the fit parameter.  The total effective electric field in a given point of space is the sum over all atomic effective fields which is converted to a polarization using the Langevin function.  Again, Newton's method is used to iteratively converge to the minimum of $\Upsilon_{EF}$.

Using a goodness of fit function related to the underlying statistics of the measurement gives the most reliable set of parameters.  In fact, artifacts in the fit parameters can be found if a typical Gaussian $\chi^2$ method is used, especially for small distances in the radial distribution functions.  It is also be noted that the parameters associated with atoms with identical underlying force field parameters are fit to a single function.  The results of these fitting procedures are found in Figure SI3 for the ADP system.


\section{Methods}
\label{methods}

Explicit solvent molecular dynamics simulations along with the Generalized Born (GB) and constant CDD simulations are performed with the AMBER 18 software package.\cite{Amber2018}  The GB model used is the GBneck model using the modified Bondi radii for the atoms.\cite{Mongan:2007}  The implicit solvent models use an external dielectric constant of 2.3473, as measured in the from the bulk chloroform model.\cite{Fox:1998, Cieplak:2001} The custom-made single ion solutes are given Lennard-Jones parameters of $\epsilon=0.152$ kcal/mol and $r_{\text{min}}=7$ \AA.  The parameters for alanine dipeptide are from the ff14SB force field.\cite{Maier:2015}  Simulations were run in the NPT ensemble at 298 K maintained using a Langevin thermostat with a collision frequency of 2 ps$^{-1}$ and a pressure of 1 bar maintained using a Berendsen barostat. A direct interaction cut off of 12 \AA\ is used with particle mesh Ewald being used to account for long range electrostatic forces.  SHAKE is employed to allow the use of a 2 fs integration time step. 
\begin{figure*}[t]
\includegraphics[width=\textwidth]{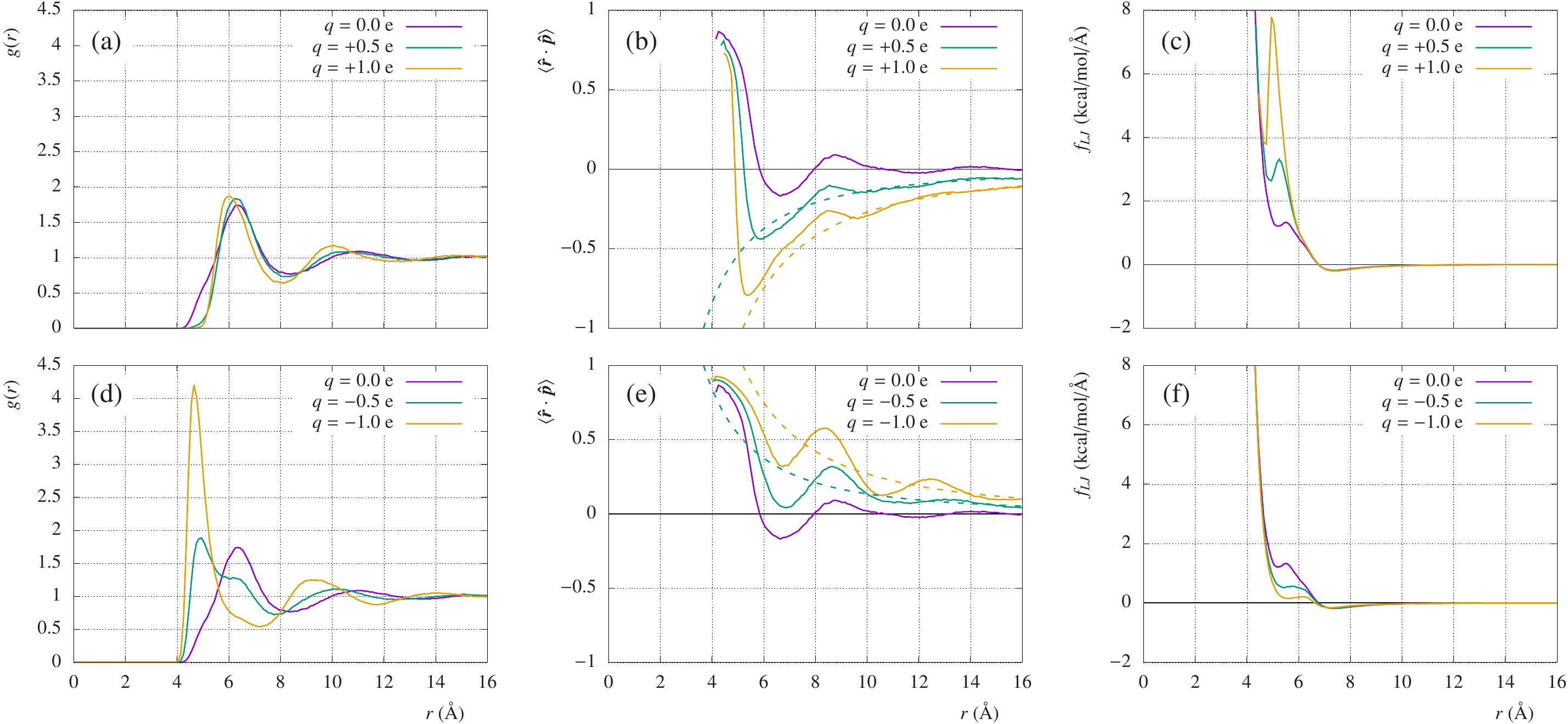}
\caption{\label{Lennard-Jones_param} IS-SPA parameters for solvation from explicit solvent simulations of neutral and charged Lennard-Jones spheres in chloroform. The top row shows behavior around the neutral and positive solute and the bottom row depicts behavior around the neutral and negative solute. (a,d) The solute--solvent radial distribution function, (b,e) The solvent polarization as a function of separation and (c,f) the mean solute--solvent Lennard-Jones force. Dashed lines in (b) and (e) are the constant density dielectric results.}%
\end{figure*}

The IS-SPA simulations are performed using our own Fortran code.  These simulations maintain the NVT ensemble with an Andersen thermostat with a collision frequency of 16.67 ps$^{-1}$.   The  mass of the hydrogen atoms is set to 12 u to reduce the frequency of those bonds and allow the use of a 2 fs integration time step.  No cutoff is used in the IS-SPA simulations and 100 MC points per atom are used for estimating the integral in \eq{exact} within a distance of 12 \AA\ of all solute atoms.

Additional simulation details are provided in the Simulation Methods section of the Supplementary Material.  These details include the umbrella sampling protocols and the magnitudes of the multipole moments of chloroform used in the IS-SPA simulations.  

\section{Results and Discussion}
\label{RandD}
\subsection{Charged Lennard-Jones Solutes in Chloroform}
The solvation and dimerization of neutral and charged Lennard-Jones spheres are used as test cases for the IS-SPA procedure.  For the ions, point charges are placed at the center of the Lennard-Jones spheres with $|q|=0$, 0.5, and 1.0 e.  The solute--solvent radial distribution function, polarization of the solvent around the solute, and mean Lennard-Jones solvent--solute force are measured from explicit solvent simulation of each of the five Lennard-Jones spheres independently.   These parameters are then used to numerically integrate the IS-SPA equations for the dimerization of oppositely charged Lennard-Jones species.

The solute--chloroform radial distribution functions for the five different Lennard-Jones spheres as measured from explicit solvent simulation are shown in \fig{Lennard-Jones_param} (a) and (d).  The $q=0$ e, $q=0.5$ e, and $q=1.0$ e curves depicted in \fig{Lennard-Jones_param}(a) all demonstrate a first solvation shell peak at $r=6$ \AA ~with $g < 2$, followed by an interstitial region, and a slight second solvation shell before achieving bulk density. The only major discrepancy between these three species is the shoulder in the $q=0$ e curve (purple curve) at $r=4.5$ \AA.  These results indicate that the chloroform molecule packs similarly around neutral and positively charged species.  
The $q=0$ e, $q=-0.5$ e, and $q=-1.0$ e curves depicted in \fig{Lennard-Jones_param}(d), however, demonstrate significant differences in solvation structure.  The $q=-0.5$ e species (green curve) has its first solvation shell peak at $r=5$ \AA ~followed by a shoulder at $r=6$ \AA ~before the chloroform starts to behave similarly to that found in around the $q=0$ e species.  The $q=-1.0$ e (yellow curve) $g(r)$ demonstrates even more dramatic discrepancy to the neutral with $g> 4$ around 5 \AA.   The discrepancy in radial packing around positive and negative ions can be attributed to the asymmetric charge distribution in the chloroform model.  The asymmetry of charge solvation is something that is also observed in water and any asymmetric polar solvent.\cite{Lake:2020b}  This asymmetry is captured in IS-SPA since the $g(r)$'s plotted in \fig{Lennard-Jones_param} (a) and (d) are used directly as parameters in the model. 
\begin{figure*}[t]
\includegraphics[width=\textwidth]{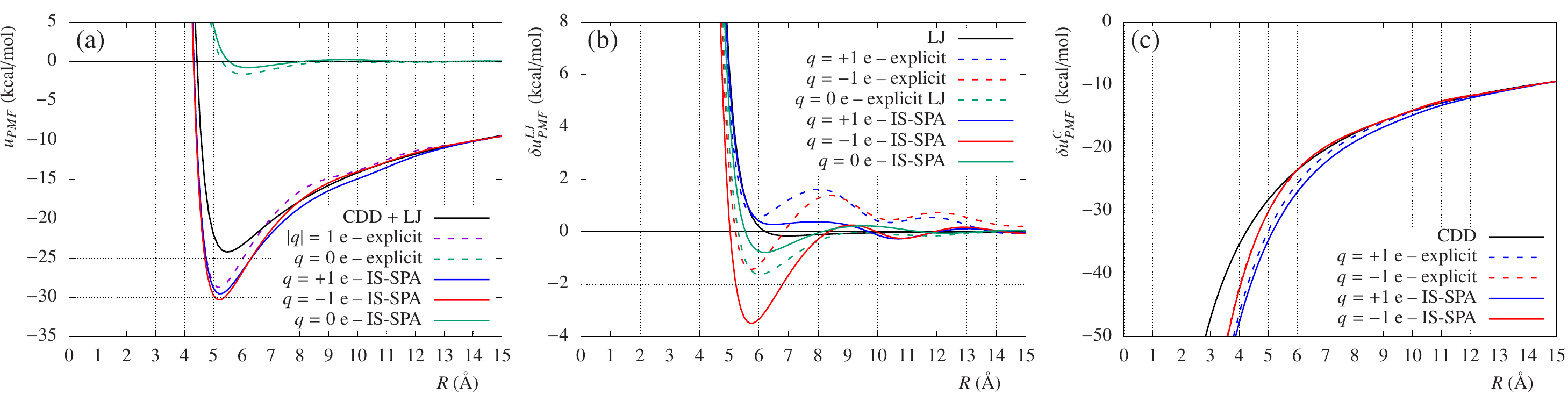}
\caption{\label{Lennard-Jones_pmf} The potential of mean force (PMF) of dimerization of two Lennard-Jones spheres in chloroform. The red and blue IS-SPA curves represent the integration of the positive and negative ion forces respectively. (a) The total PMF of dimerization for neutral and oppositely charged ions in various models of chloroform. (b) The Lennard-Jones component of the PMF for these same models and (c) the Coulombic component of the PMF for these same models.}%
\end{figure*}

The polarization of chloroform around the solute is an additional parameter that is measured from explicit solvent simulation and utilized in the IS-SPA equations.  From simulation of each solute, the average orientation of the dipole moment of chloroform, $\bm{\hat{p}}$, relative to the solute--solvent separation vector, $\bm{\hat{r}}$, is measured as a function of solute--solvent separation distance.  These functions are plotted for the five charged Lennard-Jones spheres in \fig{Lennard-Jones_param} (b) and (e).  Given the orientation of the dipole moment in chloroform depicted in \fig{chloroform_schematic}, a value of $\langle\bm{\hat{r}}\cdot\bm{\hat{p}}\rangle=1$ indicates that the hydrogen of the chloroform is pointing towards the solute and a value of $\langle\bm{\hat{r}}\cdot\bm{\hat{p}}\rangle=-1$ indicates that the hydrogen of the chloroform is pointing away from the solute.  Chloroform shows little preferential orientation around the neutral solute (purple curves in \fig{Lennard-Jones_param} (b) and (e)) until it comes within $8$ \AA ~of the solute.  At distances $6<r<8$ \AA, the hydrogen orients away from the solute until $r<6$ \AA ~at which point the hydrogen orients toward the solute.  This feature is due to definition of the molecular center being closer to the hydrogen atom and thus the chloroform must orient in this manner to be able to pack close to the solute.  This orientation is also preferred around positively charged solutes at short distances despite the Coulombic repulsion between the positively charged hydrogen and solute.  An orientation with the hydrogen pointing away from the solute is observed for the positively charged solutes at distances $r>5$ \AA, as expected.  The decay of the solvent orientation behavior agrees with the CDD result (dashed lines, dielectric of 2.3473) at distance $r>9$ \AA.  The orientation of the dipole plateaus to a magnitude of 1 at around $5$ \AA ~that is not predicted by the CDD result but is the result of having a physical dipole.  Regardless of the exact origin of the oscillations in the polarization, these aspects of the individual solutes' solvation are fed directly into the IS-SPA model.

The final component necessary to parameterize IS-SPA for Lennard-Jones sphere solutes are the force functions.  The electrostatic force functions are taken as the multipole expansion through the octupole moment as measured from the minimum energy structure of the solvent molecule, as presented in the Supplementary Information.  The Lennard-Jones force function is truncated at the radial component and is tabulated as the average solvent--solute Lennard-Jones force dotted into the unit separation vector as a function of separation as measured in simulation.  These are plotted for the five Lennard-Jones solutes in \fig{Lennard-Jones_param}(c) and (f) for the five different solute species.  The shoulders and peaks observed at $r\approx 5$ \AA ~demonstrate that these functions are not strictly Lennard-Jones.  Like Lennard-Jones functions, these functions decay to zero at large distances due to the short ranged nature of the interaction.

Using these parameters, IS-SPA reproduces the dimer potential PMF for charged and uncharged Lennard-Jones spheres with high fidelity.  For this analysis, we look at the PMF of dimerization of the neutral pair and the $|q|=1$ e pair, plotted in \fig{Lennard-Jones_pmf}(a). The green curves are for the neutral pair with the explicit curve in dashed lines and the IS-SPA curve in solid.  The neutral species has a minimum in the free energy of $-1.63$ kcal/mol and $-0.79$ kcal/mol from explicit and IS-SPA respectively at a separation distance of 6 \AA.  Both IS-SPA and explicit solvent demonstrate a slight desolvation barrier at 9.5 \AA~ with minimal correlations beyond this distance.  The charged Lennard-Jones spheres demonstrate a large propensity to dimerize with the explicit (dashed purple) and IS-SPA (solid blue and red) binding free energies being close to $30$ kcal/mol.  The CDD result under predicts the stabilization of the contact pair by approximately $5$ kcal/mol.  The IS-SPA free energy curve for the oppositely charged solutes shows high fidelity with explicit over the entire domain plotted, lending strong support that IS-SPA is capturing the correct physics of ion solvation in chloroform.  
\begin{figure*}
\includegraphics[width=\textwidth]{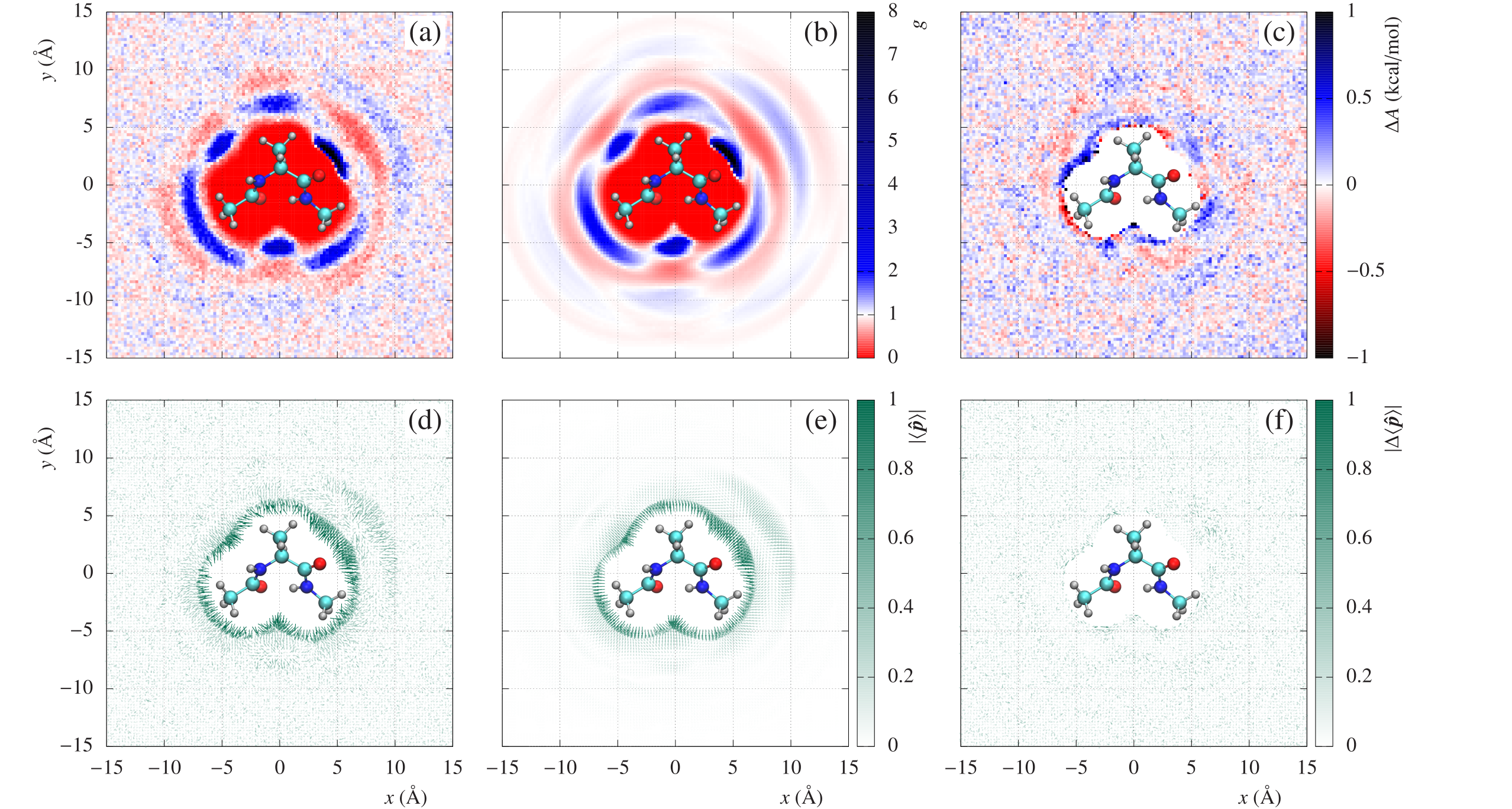}
\caption{\label{ADP_monomer_slice} Chloroform densities and polarization around a single configuration of alanine dipeptide. (a) Explicit solvent density, (b) IS-SPA density fit to explicit density, and (c) free energy difference between explicit and IS-SPA chloroform densities. (d) Explicit chloroform polarization, (e) IS-SPA polarization fit to explicit, and (f) difference in explicit and IS-SPA polarizations.}%
\end{figure*}

The Lennard-Jones and Coulombic components of the free energy are computed separately in the IS-SPA framework and can be assessed from explicit solvent using a free energy decomposition scheme.\cite{Anderson:2018} \fig{Lennard-Jones_pmf}(b) shows the Lennard-Jones component of the PMF and \fig{Lennard-Jones_pmf}(c) shows the Coulombic component of the PMF. The oscillatory nature of the Lennard-Jones component is qualitatively captured by IS-SPA; both explicit (dashed lines) and IS-SPA (solid lines) demonstrate an attractive component for the Lennard-Jones component to the PMF of the negative ion (red) and a repulsive component to the PMF of the positive ion (blue). The discrepancies between IS-SPA and explicit solvent Lennard-Jones components of the PMF are outweighed by the dominant Coulombic contribution depicted in \fig{Lennard-Jones_pmf}(c). Of particular importance is how the positive and negative ions have slightly different Coulombic contributions, which is captured by IS-SPA. Additionally, the CDD result agrees at large distance, as expected, but deviates from the explicit and IS-SPA results of both ions at distances shorter than $5$ \AA.

The Lennard-Jones and Coulombic components of the PMF need not be identical for the positive and negative ion, as they are not state functions, but the sum of the two must be equivalent.  To demonstrate this, we integrate the mean forces from IS-SPA on the positive and negative ions separately and plotted them in solid red and blue curves in \fig{Lennard-Jones_pmf}(a).  Despite the different parameters for the cation and anion, IS-SPA produces nearly identical PMFs for both species during their dimerization.  That IS-SPA accurately reproduces the explicit PMF for both the positive and negative ions despite having no closure relation to restrict such a equivalence is an important test of the method.

\subsection{Alanine Dipeptide in Chloroform}
Alanine dipeptide (ADP) is chosen as a model molecular system since it is a well studied model system for solvent models\cite{Ishizuka:2010,Wu:2011} as well as enhanced sampling methods.\cite{}  ADP has been mostly studied in aqueous environment but there have been studies of how solvent, including chloroform, affects the observed configurations.\cite{Drozdov:2004,Cai:2015,RubioMartinez:2017} As in these previous studies, we are concerned with the monomer configurations observed in a solvent of chloroform as quantified by the $\phi$ and $\psi$ backbone dihedrals.  
\begin{figure*}
\includegraphics[width=\textwidth]{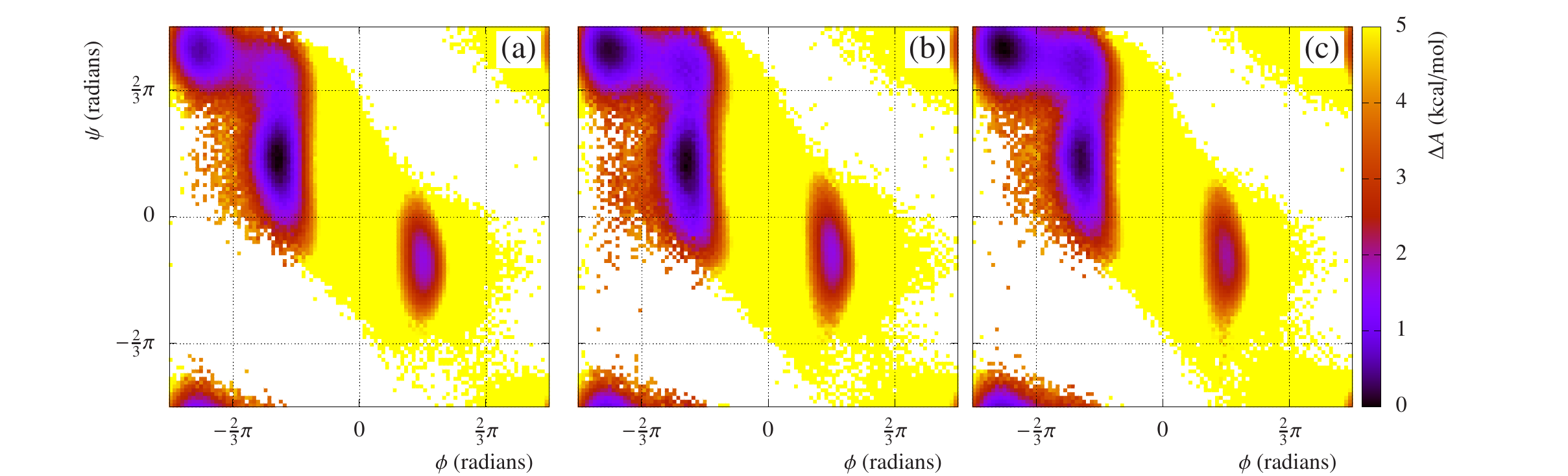}
\caption{\label{ADP_ramachandran} Ramachandran plots of alanine dipeptide in (a) vacuum, (b) explicit chloroform, and (c) IS-SPA chloroform.}%
\end{figure*}

The solvent density around ADP is used to fit atom-based radial density functions.  For molecular systems, a particular solute configuration is chosen and explicit solvent simulation is run to measure the solvent distribution around the solute.  The chosen solute configuration and a 2D slice of the 3D measured explicit solvent density are shown in \fig{ADP_monomer_slice}(a).  In this plot, white pixels indicate bulk density, red indicates below bulk density, and blue indicates above bulk density.  The excluded volume of the solute molecule is evident as the contiguous red area in the center of the figure.  The blue to black regions just outside the excluded volume are the first solvation shell of the molecule.  Two subsequent rings of low and high density are evident as the solvent gets radially farther away from the solute followed by the noisy region of bulk density.  The complete 3D data set is used to fit radial atomic densities (results provided in Figure SI1) using the Poisson regression in \eq{fit_g}.  The resulting SPA fit densities are shown in \fig{ADP_monomer_slice}(b) and a quantitative comparison of the relative free energies is shown in \fig{ADP_monomer_slice}(c).  Overall, agreement is observed in comparison with the explicit density.  While there are discrepancies between the SPA and explicit, in particular in the immediate vicinity of the solute molecule, the free energy differences are within the thermal energy of the system.  Thus, we conclude that SPA and the fitting procedure performed here are sufficient at reproducing the solvent density around a given configuration of the solute.  

Similarly, the solvent polarization around ADP is used to fit atom-based radial electric mean fields.  A 2D slice of the 3D chloroform polarization around ADP measured from explicit solvent simulation is presented as a vector field in \fig{ADP_monomer_slice}(d).  These are fit to atomic radial mean fields using the regression in \eq{fit_p}, resulting in the polarization depicted in \fig{ADP_monomer_slice}(e).  A difference map is present in \fig{ADP_monomer_slice}(f) with little quantitative difference observed.  Thus, we conclude that the SPA and fitting procedure performed here are sufficient to capture the polarization of chloroform around this single orientation of ADP.  

The atom-based parameters for density and polarization used to generate \fig{ADP_monomer_slice}(b) and (d), in combination with the force functions, can be used to compute the mean solvent force as a function of ADP internal coordinates.  Unlike the results for the Lennard-Jones spheres, the molecular systems necessitate a sampling protocol for which we developed our own IS-SPA simulation code. The results from the IS-SPA simulation are compared to vacuum and explicit chloroform simulations by looking at the relative free energy from each simulation as a function of two backbone dihedral angles, $\phi$ and $\psi$.  Also known as a Ramachandran plot, the free energy plots are depicted in \fig{ADP_ramachandran} for (a) vacuum, (b) explicit solvent, and (c) IS-SPA.  A low relative free energy is depicted in black to purple and indicates a high propensity for the simulation to populate that state; a low propensity is indicated in yellow with regions in white never being sampled.  All three systems (vacuum, explicit, and IS-SPA) have  free energy wells in four regions: C$_5$ ($\frac{2}{3}\pi < \phi < -\frac{2}{3}\pi$, $\frac{2}{3}\pi < \psi < -\frac{2}{3}\pi$), P$_\text{II}$ ($-\frac{2}{3}\pi < \phi < 0$, $\frac{2}{3}\pi < \psi < -\frac{2}{3}\pi$), C$_7^\text{eq}$ ($-\frac{2}{3}\pi < \phi < 0$, $0 < \psi < \frac{2}{3}\pi$), and C$_7^\text{axial}$ ($0 < \phi < \frac{2}{3}\pi$, $-\frac{2}{3}\pi < \psi < 0$).  There are discrepancies, however, in the relative populations of each of these states, as quantified in \tab{ADP_ramachandran_population}.  The dominant state in vacuum is found to be C$_7^\text{eq}$ with a probability of 64.4\% which is depleted in both explicit (47.9\%) and IS-SPA (36.7\%) models of chloroform.  The dominant increase due to solvation is seen in the C$_5$ populations of both explicit (32.20\%) and IS-SPA (41.18\%) as compared to vacuum (22.38\%).  Based on the probability of these four states, IS-SPA captures the impact of chloroform solvation on the ADP monomer.     

 \begin{table*}
 \caption{\label{ADP_ramachandran_population} Percent population of monomer dihedral states in alanine dipeptide in different solvent models.  The number in parentheses is the error in the last digit(s).}
 \begin{tabular}{l | d{3.8}  d{3.8}  d{3.8}  d{3.8}  d{3.8} }
  & \multicolumn{1}{c}{Vacuum} & \multicolumn{1}{c}{Explicit} & \multicolumn{1}{c}{IS-SPA}  &\multicolumn{1}{c}{CDD} & \multicolumn{1}{c}{GB}  \\\hline\hline
C$_5$         & 22.48(13)   & 32.20(18)  & 41.18(19) & 48.5(2)    & 33.6(2)      \\
P$_\text{II}$      &  8.74(10)   & 14.01(13)  & 19.00(14) & 30.6(2)    & 30.5(2)      \\
C$_7^\text{eq}$    & 64.4(3)     & 47.9(3)    & 36.7(2)   & 13.76(15)  & 25.8(2)      \\
C$_7^\text{axial}$ &  2.883(11)  &  2.219(10) &  1.294(6) &  0.235(3)  &  0.842(6)   \\
\end{tabular}
\end{table*}

In addition to vacuum, explicit and IS-SPA simulations of ADP monomer, we also ran simulations of ADP with a CDD and GB solvent with the dielectric constant of the solvent set to 2.3473 to match the explicit chloroform.  The percent populations of the four states discussed above are provided in \tab{ADP_ramachandran_population} for CDD and GB models in addition to vacuum, explicit and IS-SPA.  The CDD model dramatically overstabilizes the C$_5$ configuration (48.5\%) and understabilizes the C$_7^\text{eq}$ configuration as compared to explicit solvent.  The GB model also understabilizes the C$_5$ configuration but overtabilizes the P$_\text{II}$ configuration as compared to explicit.  To more concisely compare these distributions, we consider the relative entropy of the Ramachandran distribution of each model (IS-SPA, CDD and GB) as compared to explicit solvent.  We include vacuum compared to explicit as a control in the values tabulated in \tab{ramachandran_relative_entropy}.  Of all the models, IS-SPA has the smallest relative entropy value of 0.072(5), indicating that it has the most similar Ramachandran distribution to explicit solvent.  Vacuum has a relative entropy to explicit solvent of 0.114(5), demonstrating the IS-SPA is more similar to explicit than vacuum is to explicit. Surprisingly, the CDD and GB results are actually worse than vacuum.  

\begin{table}
\caption{\label{ramachandran_relative_entropy}Relative entropy (unitless) between the distribution of backbone dihedral angles for explicit solvent and indicated model system.  The value in parentheses is the error in the last digit(s).}
\begin{tabular}{l | d{3.8}}
\multicolumn{1}{l|}{Model} & \multicolumn{1}{c}{$S_\text{rel}$} \\\hline\hline
Vacuum &  0.114(5)  \\
CDD    &  0.762(13) \\
GB     &  0.268(7)  \\
IS-SPA &  0.072(5)  \\
\end{tabular}
\end{table}

\subsection{Dimerization of Alanine Dipeptide in Chloroform}

\begin{figure}
\includegraphics[width=0.48\textwidth]{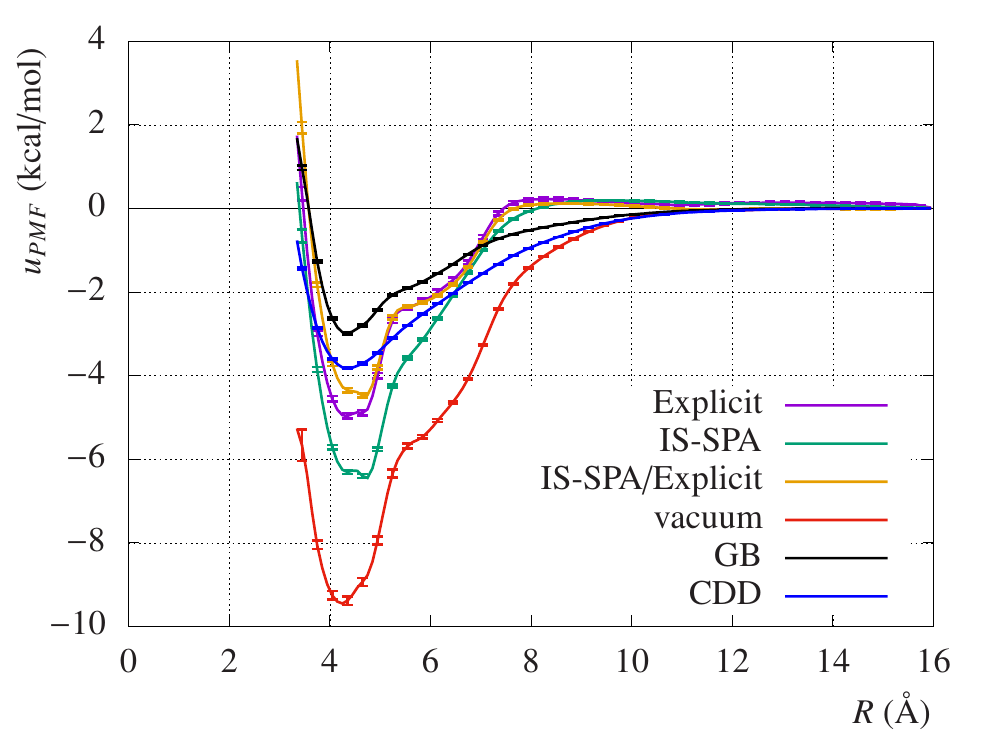}
\caption{\label{ADP_dimer_pmf} The potential of mean force (PMF) as a function of center-of-mass separation. $R$, for the dimerization of alanine dipeptide (ADP).  Five different solvation models are considered: Explicit, IS-SPA, vacuum, GB, CDD.  Additionally, the PMF is computed by integrating the mean force from IS-SPA along the Explicit solvent configurations and this is denoted IS-SPA/Explicit.}%
\end{figure}

The IS-SPA parameters developed for the monomer of ADP can also be used to simulate a dimer of ADP.  To quantitatively compare with explicit solvent, we investigate the dimerization behavior along the center-of-mass separation distance.  We utilize umbrella sampling simulations of five different models: vacuum, explicit, IS-SPA, GB and CDD.  Additionally, we compute the PMF by integrating the mean force of IS-SPA using the configurations sampled in the explicit solvent simulations and refer to this data as `IS-SPA/Explicit'.  The resulting PMFs as a function of center-of-mass separation between the monomers are shown in \fig{ADP_dimer_pmf}.  

IS-SPA correctly captures the effect of chloroform solvation on the dimerization of ADP.  This is evident in two aspects of the PMFs shown in \fig{ADP_dimer_pmf}.  The first aspect is the dimerization free energy of explicit and IS-SPA as compared to vacuum. ADP dimerization has a minimum in the PMF of 5.0 kcal/mol in explicit solvent as compared to 9.5 kcal/mol in vacuum.  This demonstrates that solvation destabilizes dimerization by 4.5 kcal/mol.  IS-SPA (green curve) destabilizes dimerization of ADP relative to gas phase by 2.7 kcal/mol.  The second aspect of the PMFs that indicate IS-SPA is correctly capturing this effect of chloroform is the position at which the PMF goes to zero.  Chloroform screens the interaction of one ADP with the other such that the attractive component of the PMF is not present until $R<8$ \AA.  This is as compared to vacuum, and the GB and CDD models, that demonstrate finite attraction between the molecules at all distances beyond contact.  
\begin{figure*}
\includegraphics[width=\textwidth]{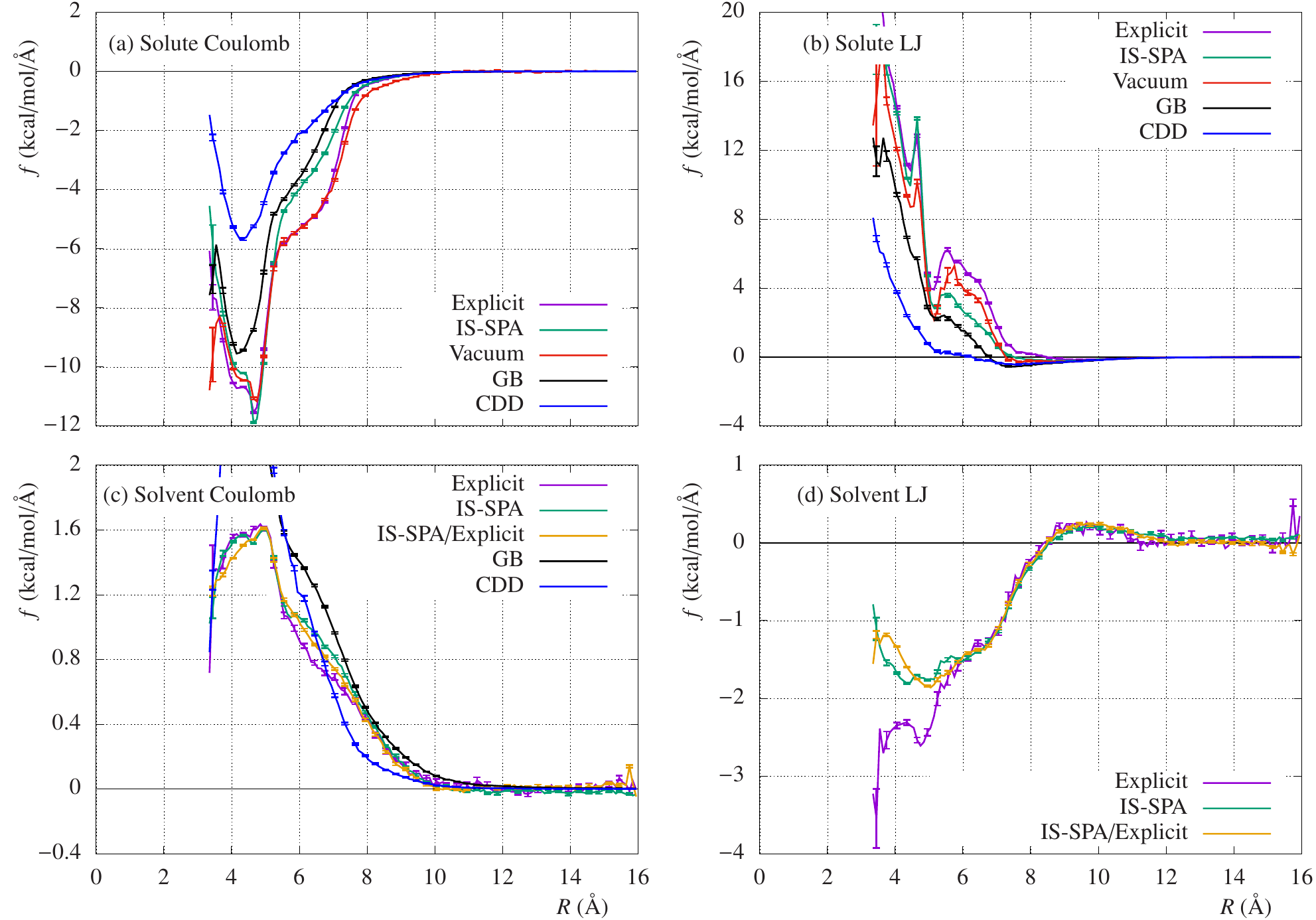}
\caption{\label{ADP_dimer_forces} Mean forces as a function of center-of-mass separation distance, $R$ for alanine dipeptide dimerization.  The total mean force is separated into: (a) solute--solute Coulomb, (b) solute--solute Lennard-Jones, (c) solute--solvent Coulomb, and (d) solute--solvent Lennard-Jones. Five different solvation models are considered: Explicit, IS-SPA, vacuum, GB, and CDD.  Additionally, the mean solute-solvent forces from the IS-SPA model computed along the explicit solvent configurations are shown under label IS-SPA/Explicit separated into Coulomb, (c), and Lennard-Jones, (d), components.}%
\end{figure*}

The quantitative discrepancy between explicit and IS-SPA stems from a disagreement between populated solute configurations between 5 and 8 \AA.  This is demonstrated in two ways.  The first piece of evidence is the impressive agreement between explicit solvent and the IS-SPA/Explicit curves in \fig{ADP_dimer_pmf}.  This curve is computed by integrating the mean IS-SPA force using the explicit solvent sampled solute configurations.   Thus, the IS-SPA curve in \fig{ADP_dimer_pmf} differs from explicit due to sampling different solute configurations.  Since the solvent force projected along the separation of the center of mass of the two molecules agree, it must be a small difference in the force acting in some other degree of freedom in the system.  The second piece of evidence is the deviation in mean solute--solute forces along the center-of-mass separation.  The mean solute--solute force is decomposed into Coulomb and Lennard-Jones components and plotted in \fig{ADP_dimer_forces}(a) and (b) respectively.  Focusing on the solute Coulomb forces, we see that Explicit (purple) and IS-SPA have finite attractive forces for $R<10$ \AA ~but that these values are discrepant for $5<R<8$ \AA.  Outside of this domain, IS-SPA and explicit have good agreement.  This suggests that IS-SPA and explicit solute configurations are almost identical except for outside of this domain.  The discrepancy in forces in this domain propagate into the PMF at the minimum upon integrating the mean force.  A similar argument can be made by investigating the solute Lennard-Jones force in \fig{ADP_dimer_forces}(b).  We note that the explicit, IS-SPA/Explicit, IS-SPA solvent--solute forces plotted in \fig{ADP_dimer_forces}(c) and (d) demonstrate less quantitative difference for $R>5$ \AA ~than the solute-solute components.  The takeaway is that IS-SPA simulations populate a different set of solute configurations than explicit in the domain $5<R<8$ \AA\ while populating the same states in the rest of the domain sampled.

IS-SPA better reproduces the explicit dimerization of the PMF than the other solvation models tested here as seen in \fig{ADP_dimer_pmf}. Unlike Explicit and IS-SPA, GB and CDD models demonstrate finite attraction out to $R>10$ \AA.  Additionally, CDD and GB understabilize the ADP contact dimer and oversimplify the curvature near the minimum.  These discrepancies can be quantified in a single parameter, $\chi^2$, defined as the integral of the squared difference between the PMFs, 
\begin{equation}
    \chi^2 = \frac{1}{T^2}\int\Delta\Delta A(R)^2 \;\text{d}R,
\end{equation}
where $\Delta \Delta A(R) = \Delta A(R)_\text{model}-\Delta A(R)_\text{explicit}$.  These values are computed for each model depicted in \fig{ADP_dimer_pmf} as compared to explicit solvent and are tabulated in \tab{ADP_dimer_FE_diff}.  The least discrepant model is IS-SPA/Explicit with a value of $\chi^2=6.0$ \AA.  With the IS-SPA simulation and the sampling of different states between $5<R<8$ \AA ~we get an IS-SPA $\chi^2=14.0$ \AA ~which is still smaller than the CDD ($\chi^2=16.6$ \AA) and GB ($\chi^2=19.5$ \AA) models.

 \begin{table}
 \caption{\label{ADP_dimer_FE_diff} Differences between PMFs shown in \fig{ADP_dimer_pmf} as quantified by the $\chi^2 = \beta^2\int\Delta\Delta A(R)^2\text{d}R$. }
 \begin{tabular}{l | d{4.8}}
  \multicolumn{1}{l|}{Model} & \multicolumn{1}{c}{$\chi^2$ (\AA)} \\\hline\hline
IS-SPA &  14.0(5)  \\
IS-SPA/Explicit    &  6.0(10) \\
Vacuum     &  207.(5)  \\
GB & 19.5(3)  \\
CDD & 16.6(11)  \\
\end{tabular}
\end{table}


\subsection{IS-SPA Algorithm Scaling}
An IS-SPA simulation follows typical all-atom molecular dynamics simulation protocols expect for the inclusion of an additional IS-SPA routine to estimate the mean solvent force on each atom.  Currently, this algorithm is performed at every step of the simulation.  For a system of $N$ solute particles and $N_\text{MC}$ MC points per particle, the algorithm is as follows:
\begin{enumerate}
    \item Generate $N\cdot N_\text{MC}$ MC points (sampled from predetermined distribution).
    \item For each MC point, loop over all atoms and use SPA to determine density and mean field at MC point.
    \item For each atom, loop over all MC points and compute Lennard-Jones and Coulomb force from MC point on atom.
\end{enumerate}
The algorithm involves two loops of size $N^2\cdot N_\text{MC}$.  The first to compute the density and mean field at each MC point and the second to push the force from that MC point onto each atom.  This is similar to what is done in GB except that we have the additional $N_\text{MC}$ points per solute atom in IS-SPA.

Considering a system of $N$ solute particles solvated with either $M$ solvent atoms per solute atom in explicit solvent or $N_\text{MC}$ MC points per solute atom in IS-SPA, we achieve the following naive (no neighbor list, no PME) performance scaling relationships for non-bonded and solvent calculations
\begin{eqnarray}
    P_{\text{IS-SPA}} \approx a\cdot N_{MC}\cdot N^2 + b\cdot N^2\\
    P_{\text{Explicit}} \approx  b\cdot (M+2)M\cdot N^2 + b\cdot N^2,
\end{eqnarray}
where $a$ and $b$ are performance coefficients in front of the IS-SPA and non-bonding loops, respectively.  We expect $a \geq 2b$ since IS-SPA involves two loops over all-pairs.  In practice, we get $b\approx 0.27a$ in our IS-SPA code due to the additional algebraic steps to compute the solvent forces.  Notice that IS-SPA does not have an $N_\text{MC}^2$ term in the expected scaling relationship since each MC point is independent of the other.  From these scaling relationships, it is apparent that IS-SPA will perform better than explicit when $N_\text{MC} < \frac{b}{a}M(M+2)$.  If we set $N_\text{MC}=100$ as used in the current work and $\frac{b}{a}=0.27$, we find that IS-SPA is expected to perform better than explicit for ADP concentrations of lower than 150 mM in chloroform.    This is significantly higher than the 54 mM high experimental concentrations of diphenylalanine in self-assembly experiments.\cite{Yan:2010}

\section{Conclusions}

In the current work, we adapt the previous IS-SPA model to accurately account for polar solutes in a polar, non-spherical solvent.  We consider the solvent to be a dipole orienting in a superposition of mean fields emanating from each solute atom.  These mean fields and radial solute--solvent densities are determined from a simulation of the monomer in explicit solvent.  Combined with a long-ranged electrostatic term and solvent--solute Coulombic forces through the octupole term, IS-SPA simulations can be performed in a procedure analogous to our previous non-polar version.

Using this model, it is demonstrated that polar solute solvation and association is accurately captured.  As a test case, the dimerization of charged Lennard-Jones spheres in chloroform using IS-SPA is found to be in high fidelity with explicit solvent simulations.  It should also be noted that the asymmetry of charge solvation is built into the parameters fed into IS-SPA.  Thus, IS-SPA captures the asymmetry of charge solvation in a polar solute such as chloroform.  More importantly, the asymmetry of the charge solvation still amounts to the same PMF on the anion and cation for the dimerization of the opposite charges.  This behavior is not guaranteed due to the lack of closure of the SPA.  

These additions to IS-SPA also accurately capture the solvation behavior of chloroform around alanine dipeptide.  The Ramachandran plot of the monomer is well replicated by the model, with IS-SPA netting the lowest relative entropy to explicit solvent out of the three solvation models tested.  The dimerization of ADP is also well captured by IS-SPA with the lowest integrated free energy difference relative to explicit solvent.  Here there is still room for improvement.  The quantitative discrepancy in the dimer PMFs between $5$ and $8$ \AA\ stem from subtle inaccuracies in the solvent force.  We hypothesize this is mainly due to the Lennard-Jones force since we see that this force is less quantitative in the Lennard-Jones spheres.  We will pursue a variety of ways to improve this including accounting for the non-radial component of the Lennard-Jones force from chloroform. 

Finally, using the na{\"i}ve performance scaling behavior we determined that the current IS-SPA model should outperform explicit solvent simulations at 150 mM peptide solute concentration.  This predicted behavior does not account for neighbor lists or long-ranged electrostatic corrections that will impact performance.  Next steps in the development of the method for broad use will be to determine how to implement cut offs, and thus the ability to use neighbor lists, while still accurately accounting for the solvent forces.  

\section*{SUPPLEMENTARY MATERIAL} 
Supporting information is provided including: simulation methods, multipole moments of chloroform, equations for the electrostatic solvent forces in IS-SPA, and example atomic parameters for alanine dipeptide.

\section*{AUTHORS' CONTRIBUTIONS} 

P.T. Lake and M.A. Mattson contributed equally to this work.

\begin{acknowledgments}
We gratefully acknowledge the Army Research Office for financial support (ARO Fund number W911NF-17-1-0383). This work used the Extreme Science and Engineering Discovery Environment (XSEDE), which is supported by National Science Foundation grant number ACI-1548562. We would specifically like to acknowledge the San Diego Supercomputer Center Comet used under XSEDE allocation CHE160008 awarded to MM. 
\end{acknowledgments}

\section*{DATA AVAILABILITY} 
The data that support the findings of this study are available from the corresponding author upon request.  All code used to parameterize and run IS-SPA for this paper are provided in the GitHub repository: \url{https://github.com/mccullaghlab/IS-SPAQ.git}.

\bibliography{isspa_chloroform_refs}

\begin{thebibliography}{33}%
\makeatletter
\providecommand \@ifxundefined [1]{%
 \@ifx{#1\undefined}
}%
\providecommand \@ifnum [1]{%
 \ifnum #1\expandafter \@firstoftwo
 \else \expandafter \@secondoftwo
 \fi
}%
\providecommand \@ifx [1]{%
 \ifx #1\expandafter \@firstoftwo
 \else \expandafter \@secondoftwo
 \fi
}%
\providecommand \natexlab [1]{#1}%
\providecommand \enquote  [1]{``#1''}%
\providecommand \bibnamefont  [1]{#1}%
\providecommand \bibfnamefont [1]{#1}%
\providecommand \citenamefont [1]{#1}%
\providecommand \href@noop [0]{\@secondoftwo}%
\providecommand \href [0]{\begingroup \@sanitize@url \@href}%
\providecommand \@href[1]{\@@startlink{#1}\@@href}%
\providecommand \@@href[1]{\endgroup#1\@@endlink}%
\providecommand \@sanitize@url [0]{\catcode `\\12\catcode `\$12\catcode
  `\&12\catcode `\#12\catcode `\^12\catcode `\_12\catcode `\%12\relax}%
\providecommand \@@startlink[1]{}%
\providecommand \@@endlink[0]{}%
\providecommand \url  [0]{\begingroup\@sanitize@url \@url }%
\providecommand \@url [1]{\endgroup\@href {#1}{\urlprefix }}%
\providecommand \urlprefix  [0]{URL }%
\providecommand \Eprint [0]{\href }%
\providecommand \doibase [0]{http://dx.doi.org/}%
\providecommand \selectlanguage [0]{\@gobble}%
\providecommand \bibinfo  [0]{\@secondoftwo}%
\providecommand \bibfield  [0]{\@secondoftwo}%
\providecommand \translation [1]{[#1]}%
\providecommand \BibitemOpen [0]{}%
\providecommand \bibitemStop [0]{}%
\providecommand \bibitemNoStop [0]{.\EOS\space}%
\providecommand \EOS [0]{\spacefactor3000\relax}%
\providecommand \BibitemShut  [1]{\csname bibitem#1\endcsname}%
\let\auto@bib@innerbib\@empty
\bibitem [{\citenamefont {Shen}\ and\ \citenamefont
  {Murphy}(1995)}]{Shen:1995}%
  \BibitemOpen
  \bibfield  {author} {\bibinfo {author} {\bibfnamefont {C.~L.}\ \bibnamefont
  {Shen}}\ and\ \bibinfo {author} {\bibfnamefont {R.~M.}\ \bibnamefont
  {Murphy}},\ }\href@noop {} {\bibfield  {journal} {\bibinfo  {journal}
  {Biophys. J.}\ }\textbf {\bibinfo {volume} {69}},\ \bibinfo {pages} {640}
  (\bibinfo {year} {1995})}\BibitemShut {NoStop}%
\bibitem [{\citenamefont {McCullagh}\ \emph {et~al.}(2008)\citenamefont
  {McCullagh}, \citenamefont {Prytkova}, \citenamefont {Tonzani}, \citenamefont
  {Winter},\ and\ \citenamefont {Schatz}}]{McCullagh:2008}%
  \BibitemOpen
  \bibfield  {author} {\bibinfo {author} {\bibfnamefont {M.}~\bibnamefont
  {McCullagh}}, \bibinfo {author} {\bibfnamefont {T.}~\bibnamefont {Prytkova}},
  \bibinfo {author} {\bibfnamefont {S.}~\bibnamefont {Tonzani}}, \bibinfo
  {author} {\bibfnamefont {N.~D.}\ \bibnamefont {Winter}}, \ and\ \bibinfo
  {author} {\bibfnamefont {G.~C.}\ \bibnamefont {Schatz}},\ }\href {\doibase
  10.1021/jp803192u} {\bibfield  {journal} {\bibinfo  {journal} {J. Phys. Chem.
  B}\ }\textbf {\bibinfo {volume} {112}},\ \bibinfo {pages} {10388} (\bibinfo
  {year} {2008})}\BibitemShut {NoStop}%
\bibitem [{\citenamefont {Lee}, \citenamefont {Stupp},\ and\ \citenamefont
  {Schatz}(2011)}]{Lee:2011}%
  \BibitemOpen
  \bibfield  {author} {\bibinfo {author} {\bibfnamefont {O.-S.}\ \bibnamefont
  {Lee}}, \bibinfo {author} {\bibfnamefont {S.~I.}\ \bibnamefont {Stupp}}, \
  and\ \bibinfo {author} {\bibfnamefont {G.~C.}\ \bibnamefont {Schatz}},\
  }\href {\doibase 10.1021/ja110966y} {\bibfield  {journal} {\bibinfo
  {journal} {J. Am. Chem. Soc.}\ }\textbf {\bibinfo {volume} {133}},\ \bibinfo
  {pages} {3677} (\bibinfo {year} {2011})}\BibitemShut {NoStop}%
\bibitem [{\citenamefont {Anderson}, \citenamefont {Lake},\ and\ \citenamefont
  {McCullagh}(2018)}]{Anderson:2018}%
  \BibitemOpen
  \bibfield  {author} {\bibinfo {author} {\bibfnamefont {J.}~\bibnamefont
  {Anderson}}, \bibinfo {author} {\bibfnamefont {P.~T.}\ \bibnamefont {Lake}},
  \ and\ \bibinfo {author} {\bibfnamefont {M.}~\bibnamefont {McCullagh}},\
  }\href {\doibase 10.1021/acs.jpcb.8b10335} {\bibfield  {journal} {\bibinfo
  {journal} {J. Phys. Chem. B}\ }\textbf {\bibinfo {volume} {122}},\ \bibinfo
  {pages} {12331} (\bibinfo {year} {2018})}\BibitemShut {NoStop}%
\bibitem [{\citenamefont {Weber}\ and\ \citenamefont
  {McCullagh}(2020)}]{Weber:2020}%
  \BibitemOpen
  \bibfield  {author} {\bibinfo {author} {\bibfnamefont {R.}~\bibnamefont
  {Weber}}\ and\ \bibinfo {author} {\bibfnamefont {M.}~\bibnamefont
  {McCullagh}},\ }\href {\doibase 10.1021/acs.jpcb.0c00048} {\bibfield
  {journal} {\bibinfo  {journal} {J. Phys. Chem. B}\ }\textbf {\bibinfo
  {volume} {124}},\ \bibinfo {pages} {1723} (\bibinfo {year}
  {2020})}\BibitemShut {NoStop}%
\bibitem [{\citenamefont {Yan}, \citenamefont {Zhu},\ and\ \citenamefont
  {Li}(2010)}]{Yan:2010}%
  \BibitemOpen
  \bibfield  {author} {\bibinfo {author} {\bibfnamefont {X.}~\bibnamefont
  {Yan}}, \bibinfo {author} {\bibfnamefont {P.}~\bibnamefont {Zhu}}, \ and\
  \bibinfo {author} {\bibfnamefont {J.}~\bibnamefont {Li}},\ }\href {\doibase
  10.1039/b915765b} {\bibfield  {journal} {\bibinfo  {journal} {Chem. Soc.
  Rev.}\ }\textbf {\bibinfo {volume} {39}},\ \bibinfo {pages} {1877} (\bibinfo
  {year} {2010})}\BibitemShut {NoStop}%
\bibitem [{\citenamefont {Mansbach}\ and\ \citenamefont
  {Ferguson}(2017)}]{Mansbach:2017}%
  \BibitemOpen
  \bibfield  {author} {\bibinfo {author} {\bibfnamefont {R.~A.}\ \bibnamefont
  {Mansbach}}\ and\ \bibinfo {author} {\bibfnamefont {A.~L.}\ \bibnamefont
  {Ferguson}},\ }\href {\doibase 10.1021/acs.jpcb.6b10165} {\bibfield
  {journal} {\bibinfo  {journal} {J. Phys. Chem. B}\ }\textbf {\bibinfo
  {volume} {121}},\ \bibinfo {pages} {1684} (\bibinfo {year}
  {2017})}\BibitemShut {NoStop}%
\bibitem [{\citenamefont {Frederix}\ \emph {et~al.}(2011)\citenamefont
  {Frederix}, \citenamefont {Ulijn}, \citenamefont {Hunt},\ and\ \citenamefont
  {Tuttle}}]{Frederix:2011}%
  \BibitemOpen
  \bibfield  {author} {\bibinfo {author} {\bibfnamefont {P.~W. J.~M.}\
  \bibnamefont {Frederix}}, \bibinfo {author} {\bibfnamefont {R.~V.}\
  \bibnamefont {Ulijn}}, \bibinfo {author} {\bibfnamefont {N.~T.}\ \bibnamefont
  {Hunt}}, \ and\ \bibinfo {author} {\bibfnamefont {T.}~\bibnamefont
  {Tuttle}},\ }\href {\doibase 10.1021/jz2010573} {\bibfield  {journal}
  {\bibinfo  {journal} {J. Phys. Chem. Lett.}\ }\textbf {\bibinfo {volume}
  {2}},\ \bibinfo {pages} {2380} (\bibinfo {year} {2011})}\BibitemShut
  {NoStop}%
\bibitem [{\citenamefont {Frederix}\ \emph {et~al.}(2015)\citenamefont
  {Frederix}, \citenamefont {Scott}, \citenamefont {Abul-Haija}, \citenamefont
  {Kalafatovic}, \citenamefont {Pappas}, \citenamefont {Javid}, \citenamefont
  {Hunt}, \citenamefont {Ulijn},\ and\ \citenamefont {Tuttle}}]{Frederix:2015}%
  \BibitemOpen
  \bibfield  {author} {\bibinfo {author} {\bibfnamefont {P.~W. J.~M.}\
  \bibnamefont {Frederix}}, \bibinfo {author} {\bibfnamefont {G.~G.}\
  \bibnamefont {Scott}}, \bibinfo {author} {\bibfnamefont {Y.~M.}\ \bibnamefont
  {Abul-Haija}}, \bibinfo {author} {\bibfnamefont {D.}~\bibnamefont
  {Kalafatovic}}, \bibinfo {author} {\bibfnamefont {C.~G.}\ \bibnamefont
  {Pappas}}, \bibinfo {author} {\bibfnamefont {N.}~\bibnamefont {Javid}},
  \bibinfo {author} {\bibfnamefont {N.~T.}\ \bibnamefont {Hunt}}, \bibinfo
  {author} {\bibfnamefont {R.~V.}\ \bibnamefont {Ulijn}}, \ and\ \bibinfo
  {author} {\bibfnamefont {T.}~\bibnamefont {Tuttle}},\ }\href {\doibase
  10.1038/nchem.2122} {\bibfield  {journal} {\bibinfo  {journal} {Nature Chem}\
  }\textbf {\bibinfo {volume} {7}},\ \bibinfo {pages} {1} (\bibinfo {year}
  {2015})}\BibitemShut {NoStop}%
\bibitem [{\citenamefont {Tuttle}(2015)}]{Tuttle:2015}%
  \BibitemOpen
  \bibfield  {author} {\bibinfo {author} {\bibfnamefont {T.}~\bibnamefont
  {Tuttle}},\ }\href {\doibase 10.1002/ijch.201400188} {\bibfield  {journal}
  {\bibinfo  {journal} {Can. J. Chem. Eng.}\ }\textbf {\bibinfo {volume}
  {55}},\ \bibinfo {pages} {724} (\bibinfo {year} {2015})}\BibitemShut
  {NoStop}%
\bibitem [{\citenamefont {Sanyal}\ and\ \citenamefont
  {Shell}(2018)}]{Sanyal:2018}%
  \BibitemOpen
  \bibfield  {author} {\bibinfo {author} {\bibfnamefont {T.}~\bibnamefont
  {Sanyal}}\ and\ \bibinfo {author} {\bibfnamefont {M.~S.}\ \bibnamefont
  {Shell}},\ }\href {\doibase 10.1021/acs.jpcb.7b12446} {\bibfield  {journal}
  {\bibinfo  {journal} {J. Phys. Chem. B}\ }\textbf {\bibinfo {volume} {122}},\
  \bibinfo {pages} {5678} (\bibinfo {year} {2018})}\BibitemShut {NoStop}%
\bibitem [{\citenamefont {Jin}, \citenamefont {Pak},\ and\ \citenamefont
  {Voth}(2019)}]{Jin:2019}%
  \BibitemOpen
  \bibfield  {author} {\bibinfo {author} {\bibfnamefont {J.}~\bibnamefont
  {Jin}}, \bibinfo {author} {\bibfnamefont {A.~J.}\ \bibnamefont {Pak}}, \ and\
  \bibinfo {author} {\bibfnamefont {G.~A.}\ \bibnamefont {Voth}},\ }\href
  {\doibase 10.1021/acs.jpclett.9b01228} {\bibfield  {journal} {\bibinfo
  {journal} {J. Phys. Chem. Lett.}\ }\textbf {\bibinfo {volume} {10}},\
  \bibinfo {pages} {4549} (\bibinfo {year} {2019})}\BibitemShut {NoStop}%
\bibitem [{\citenamefont {Zhao}\ \emph {et~al.}(2019)\citenamefont {Zhao},
  \citenamefont {Liao}, \citenamefont {Ma}, \citenamefont {Ferrell},
  \citenamefont {Schneebeli},\ and\ \citenamefont {Li}}]{Zhao:2019}%
  \BibitemOpen
  \bibfield  {author} {\bibinfo {author} {\bibfnamefont {X.}~\bibnamefont
  {Zhao}}, \bibinfo {author} {\bibfnamefont {C.}~\bibnamefont {Liao}}, \bibinfo
  {author} {\bibfnamefont {Y.-T.}\ \bibnamefont {Ma}}, \bibinfo {author}
  {\bibfnamefont {J.~B.}\ \bibnamefont {Ferrell}}, \bibinfo {author}
  {\bibfnamefont {S.~T.}\ \bibnamefont {Schneebeli}}, \ and\ \bibinfo {author}
  {\bibfnamefont {J.}~\bibnamefont {Li}},\ }\href {\doibase
  10.1021/acs.jctc.8b01025} {\bibfield  {journal} {\bibinfo  {journal} {J.
  Chem. Theory Comput.}\ }\textbf {\bibinfo {volume} {15}},\ \bibinfo {pages}
  {1514} (\bibinfo {year} {2019})}\BibitemShut {NoStop}%
\bibitem [{\citenamefont {Colombo}, \citenamefont {Soto},\ and\ \citenamefont
  {Gazit}(2007)}]{Colombo:2007}%
  \BibitemOpen
  \bibfield  {author} {\bibinfo {author} {\bibfnamefont {G.}~\bibnamefont
  {Colombo}}, \bibinfo {author} {\bibfnamefont {P.}~\bibnamefont {Soto}}, \
  and\ \bibinfo {author} {\bibfnamefont {E.}~\bibnamefont {Gazit}},\ }\href
  {\doibase 10.1016/j.tibtech.2007.03.004} {\bibfield  {journal} {\bibinfo
  {journal} {Trends Biotechnol.}\ }\textbf {\bibinfo {volume} {25}},\ \bibinfo
  {pages} {211} (\bibinfo {year} {2007})}\BibitemShut {NoStop}%
\bibitem [{\citenamefont {Shell}(2008)}]{Shell:2008}%
  \BibitemOpen
  \bibfield  {author} {\bibinfo {author} {\bibfnamefont {M.~S.}\ \bibnamefont
  {Shell}},\ }\href {\doibase 10.1063/1.2992060} {\bibfield  {journal}
  {\bibinfo  {journal} {J. Chem. Phys.}\ }\textbf {\bibinfo {volume} {129}},\
  \bibinfo {pages} {144108} (\bibinfo {year} {2008})}\BibitemShut {NoStop}%
\bibitem [{\citenamefont {Ishizuka}, \citenamefont {Huber},\ and\ \citenamefont
  {McCammon}(2010)}]{Ishizuka:2010}%
  \BibitemOpen
  \bibfield  {author} {\bibinfo {author} {\bibfnamefont {R.}~\bibnamefont
  {Ishizuka}}, \bibinfo {author} {\bibfnamefont {G.~A.}\ \bibnamefont {Huber}},
  \ and\ \bibinfo {author} {\bibfnamefont {J.~A.}\ \bibnamefont {McCammon}},\
  }\href {\doibase 10.1021/jz100665c} {\bibfield  {journal} {\bibinfo
  {journal} {J. Phys. Chem. Lett.}\ ,\ \bibinfo {pages} {2279}} (\bibinfo
  {year} {2010})}\BibitemShut {NoStop}%
\bibitem [{\citenamefont {Remsing}, \citenamefont {Rodgers},\ and\
  \citenamefont {Weeks}(2011)}]{Remsing:2011}%
  \BibitemOpen
  \bibfield  {author} {\bibinfo {author} {\bibfnamefont {R.~C.}\ \bibnamefont
  {Remsing}}, \bibinfo {author} {\bibfnamefont {J.~M.}\ \bibnamefont
  {Rodgers}}, \ and\ \bibinfo {author} {\bibfnamefont {J.~D.}\ \bibnamefont
  {Weeks}},\ }\href {\doibase 10.1007/s10955-011-0299-3} {\bibfield  {journal}
  {\bibinfo  {journal} {J. Stat. Phys.}\ }\textbf {\bibinfo {volume} {145}},\
  \bibinfo {pages} {313} (\bibinfo {year} {2011})}\BibitemShut {NoStop}%
\bibitem [{\citenamefont {Remsing}, \citenamefont {Liu},\ and\ \citenamefont
  {Weeks}(2016)}]{Remsing:2016}%
  \BibitemOpen
  \bibfield  {author} {\bibinfo {author} {\bibfnamefont {R.~C.}\ \bibnamefont
  {Remsing}}, \bibinfo {author} {\bibfnamefont {S.}~\bibnamefont {Liu}}, \ and\
  \bibinfo {author} {\bibfnamefont {J.~D.}\ \bibnamefont {Weeks}},\ }\href
  {\doibase 10.1073/pnas.1521570113} {\bibfield  {journal} {\bibinfo  {journal}
  {Proc. Natl. Acad. Sci. U.S.A.}\ }\textbf {\bibinfo {volume} {113}},\
  \bibinfo {pages} {2819} (\bibinfo {year} {2016})}\BibitemShut {NoStop}%
\bibitem [{\citenamefont {Gao}, \citenamefont {Remsing},\ and\ \citenamefont
  {Weeks}(2020)}]{Gao:2020}%
  \BibitemOpen
  \bibfield  {author} {\bibinfo {author} {\bibfnamefont {A.}~\bibnamefont
  {Gao}}, \bibinfo {author} {\bibfnamefont {R.~C.}\ \bibnamefont {Remsing}}, \
  and\ \bibinfo {author} {\bibfnamefont {J.~D.}\ \bibnamefont {Weeks}},\ }\href
  {\doibase 10.1073/pnas.1918981117} {\bibfield  {journal} {\bibinfo  {journal}
  {Proc. Natl. Acad. Sci. U.S.A.}\ }\textbf {\bibinfo {volume} {179}},\
  \bibinfo {pages} {201918981} (\bibinfo {year} {2020})}\BibitemShut {NoStop}%
\bibitem [{\citenamefont {Lake}\ and\ \citenamefont
  {McCullagh}(2017)}]{Lake:2017}%
  \BibitemOpen
  \bibfield  {author} {\bibinfo {author} {\bibfnamefont {P.~T.}\ \bibnamefont
  {Lake}}\ and\ \bibinfo {author} {\bibfnamefont {M.}~\bibnamefont
  {McCullagh}},\ }\href {\doibase 10.1021/acs.jctc.7b00698} {\bibfield
  {journal} {\bibinfo  {journal} {J. Chem. Theory Comput.}\ }\textbf {\bibinfo
  {volume} {13}},\ \bibinfo {pages} {5911} (\bibinfo {year}
  {2017})}\BibitemShut {NoStop}%
\bibitem [{\citenamefont {Kirkwood}(1935)}]{Kirkwood:1935}%
  \BibitemOpen
  \bibfield  {author} {\bibinfo {author} {\bibfnamefont {J.~G.}\ \bibnamefont
  {Kirkwood}},\ }\href@noop {} {\bibfield  {journal} {\bibinfo  {journal} {J.
  Chem. Phys.}\ }\textbf {\bibinfo {volume} {3}},\ \bibinfo {pages} {300}
  (\bibinfo {year} {1935})}\BibitemShut {NoStop}%
\bibitem [{\citenamefont {Pellegrini}\ and\ \citenamefont
  {Doniach}(1995)}]{Pellegrini:1995}%
  \BibitemOpen
  \bibfield  {author} {\bibinfo {author} {\bibfnamefont {M.}~\bibnamefont
  {Pellegrini}}\ and\ \bibinfo {author} {\bibfnamefont {S.}~\bibnamefont
  {Doniach}},\ }\href {\doibase 10.1063/1.470503} {\bibfield  {journal}
  {\bibinfo  {journal} {J. Chem. Phys.}\ }\textbf {\bibinfo {volume} {103}},\
  \bibinfo {pages} {2696} (\bibinfo {year} {1995})}\BibitemShut {NoStop}%
\bibitem [{\citenamefont {Pellegrini}, \citenamefont {Gronbech-Jensen},\ and\
  \citenamefont {Doniach}(1996)}]{Pellegrini:1996}%
  \BibitemOpen
  \bibfield  {author} {\bibinfo {author} {\bibfnamefont {M.}~\bibnamefont
  {Pellegrini}}, \bibinfo {author} {\bibfnamefont {N.}~\bibnamefont
  {Gronbech-Jensen}}, \ and\ \bibinfo {author} {\bibfnamefont {S.}~\bibnamefont
  {Doniach}},\ }\href {\doibase 10.1063/1.471552} {\bibfield  {journal}
  {\bibinfo  {journal} {J. Chem. Phys.}\ }\textbf {\bibinfo {volume} {104}},\
  \bibinfo {pages} {8639} (\bibinfo {year} {1996})}\BibitemShut {NoStop}%
\bibitem [{\citenamefont {Case}\ \emph {et~al.}(2018)\citenamefont {Case},
  \citenamefont {Betz}, \citenamefont {Cerutti}, \citenamefont {{Cheatham
  III}}, \citenamefont {Darden}, \citenamefont {Duke}, \citenamefont {Giese},
  \citenamefont {Gohlke}, \citenamefont {Goetz}, \citenamefont {Homeyer},
  \citenamefont {Izadi}, \citenamefont {Janowski}, \citenamefont {Kaus},
  \citenamefont {Kovalenko}, \citenamefont {Lee}, \citenamefont {LeGrand},
  \citenamefont {Li}, \citenamefont {Lin}, \citenamefont {Luchko},
  \citenamefont {Luo}, \citenamefont {Madej}, \citenamefont {Mermelstein},
  \citenamefont {Merz}, \citenamefont {Monard}, \citenamefont {Nguyen},
  \citenamefont {Nguyen}, \citenamefont {Omelyan}, \citenamefont {Onufriev},
  \citenamefont {Roe}, \citenamefont {Roitberg}, \citenamefont {Sagui},
  \citenamefont {Simmerling}, \citenamefont {Botello-Smith}, \citenamefont
  {Swails}, \citenamefont {Walker}, \citenamefont {Wang}, \citenamefont {Wolf},
  \citenamefont {Wu}, \citenamefont {Xiao},\ and\ \citenamefont
  {Kollman}}]{Amber2018}%
  \BibitemOpen
  \bibfield  {author} {\bibinfo {author} {\bibfnamefont {D.}~\bibnamefont
  {Case}}, \bibinfo {author} {\bibfnamefont {R.}~\bibnamefont {Betz}}, \bibinfo
  {author} {\bibfnamefont {D.}~\bibnamefont {Cerutti}}, \bibinfo {author}
  {\bibfnamefont {T.}~\bibnamefont {{Cheatham III}}}, \bibinfo {author}
  {\bibfnamefont {T.}~\bibnamefont {Darden}}, \bibinfo {author} {\bibfnamefont
  {R.}~\bibnamefont {Duke}}, \bibinfo {author} {\bibfnamefont {T.}~\bibnamefont
  {Giese}}, \bibinfo {author} {\bibfnamefont {H.}~\bibnamefont {Gohlke}},
  \bibinfo {author} {\bibfnamefont {A.}~\bibnamefont {Goetz}}, \bibinfo
  {author} {\bibfnamefont {N.}~\bibnamefont {Homeyer}}, \bibinfo {author}
  {\bibfnamefont {S.}~\bibnamefont {Izadi}}, \bibinfo {author} {\bibfnamefont
  {P.}~\bibnamefont {Janowski}}, \bibinfo {author} {\bibfnamefont
  {J.}~\bibnamefont {Kaus}}, \bibinfo {author} {\bibfnamefont {A.}~\bibnamefont
  {Kovalenko}}, \bibinfo {author} {\bibfnamefont {T.}~\bibnamefont {Lee}},
  \bibinfo {author} {\bibfnamefont {S.}~\bibnamefont {LeGrand}}, \bibinfo
  {author} {\bibfnamefont {P.}~\bibnamefont {Li}}, \bibinfo {author}
  {\bibfnamefont {C.}~\bibnamefont {Lin}}, \bibinfo {author} {\bibfnamefont
  {T.}~\bibnamefont {Luchko}}, \bibinfo {author} {\bibfnamefont
  {R.}~\bibnamefont {Luo}}, \bibinfo {author} {\bibfnamefont {B.}~\bibnamefont
  {Madej}}, \bibinfo {author} {\bibfnamefont {D.}~\bibnamefont {Mermelstein}},
  \bibinfo {author} {\bibfnamefont {K.}~\bibnamefont {Merz}}, \bibinfo {author}
  {\bibfnamefont {G.}~\bibnamefont {Monard}}, \bibinfo {author} {\bibfnamefont
  {H.}~\bibnamefont {Nguyen}}, \bibinfo {author} {\bibfnamefont
  {H.}~\bibnamefont {Nguyen}}, \bibinfo {author} {\bibfnamefont
  {I.}~\bibnamefont {Omelyan}}, \bibinfo {author} {\bibfnamefont
  {A.}~\bibnamefont {Onufriev}}, \bibinfo {author} {\bibfnamefont
  {D.}~\bibnamefont {Roe}}, \bibinfo {author} {\bibfnamefont {A.}~\bibnamefont
  {Roitberg}}, \bibinfo {author} {\bibfnamefont {C.}~\bibnamefont {Sagui}},
  \bibinfo {author} {\bibfnamefont {C.}~\bibnamefont {Simmerling}}, \bibinfo
  {author} {\bibfnamefont {W.}~\bibnamefont {Botello-Smith}}, \bibinfo {author}
  {\bibfnamefont {J.}~\bibnamefont {Swails}}, \bibinfo {author} {\bibfnamefont
  {R.}~\bibnamefont {Walker}}, \bibinfo {author} {\bibfnamefont
  {J.}~\bibnamefont {Wang}}, \bibinfo {author} {\bibfnamefont {R.}~\bibnamefont
  {Wolf}}, \bibinfo {author} {\bibfnamefont {X.}~\bibnamefont {Wu}}, \bibinfo
  {author} {\bibfnamefont {L.}~\bibnamefont {Xiao}}, \ and\ \bibinfo {author}
  {\bibfnamefont {P.}~\bibnamefont {Kollman}},\ }\href@noop {} {\emph {\bibinfo
  {title} {{Amber 2018}}}}\ (\bibinfo  {publisher} {University of California,
  San Francisco.},\ \bibinfo {year} {2018})\BibitemShut {NoStop}%
\bibitem [{\citenamefont {Mongan}\ \emph {et~al.}(2007)\citenamefont {Mongan},
  \citenamefont {Simmerling}, \citenamefont {A.~McCammon}, \citenamefont
  {A.~Case},\ and\ \citenamefont {Onufriev}}]{Mongan:2007}%
  \BibitemOpen
  \bibfield  {author} {\bibinfo {author} {\bibfnamefont {J.}~\bibnamefont
  {Mongan}}, \bibinfo {author} {\bibfnamefont {C.}~\bibnamefont {Simmerling}},
  \bibinfo {author} {\bibfnamefont {J.}~\bibnamefont {A.~McCammon}}, \bibinfo
  {author} {\bibfnamefont {D.}~\bibnamefont {A.~Case}}, \ and\ \bibinfo
  {author} {\bibfnamefont {A.}~\bibnamefont {Onufriev}},\ }\href@noop {}
  {\bibfield  {journal} {\bibinfo  {journal} {. Chem. Theory Comput.}\ }\textbf
  {\bibinfo {volume} {3}},\ \bibinfo {pages} {156} (\bibinfo {year}
  {2007})}\BibitemShut {NoStop}%
\bibitem [{\citenamefont {Fox}\ and\ \citenamefont {Kollman}(1998)}]{Fox:1998}%
  \BibitemOpen
  \bibfield  {author} {\bibinfo {author} {\bibfnamefont {T.}~\bibnamefont
  {Fox}}\ and\ \bibinfo {author} {\bibfnamefont {P.~A.}\ \bibnamefont
  {Kollman}},\ }\href {\doibase 10.1021/jp9717655} {\bibfield  {journal}
  {\bibinfo  {journal} {Langmuir}\ }\textbf {\bibinfo {volume} {102}},\
  \bibinfo {pages} {8070} (\bibinfo {year} {1998})}\BibitemShut {NoStop}%
\bibitem [{\citenamefont {Cieplak}, \citenamefont {Caldwell},\ and\
  \citenamefont {Kollman}(2001)}]{Cieplak:2001}%
  \BibitemOpen
  \bibfield  {author} {\bibinfo {author} {\bibfnamefont {P.}~\bibnamefont
  {Cieplak}}, \bibinfo {author} {\bibfnamefont {J.}~\bibnamefont {Caldwell}}, \
  and\ \bibinfo {author} {\bibfnamefont {P.}~\bibnamefont {Kollman}},\ }\href
  {\doibase 10.1002/jcc.1065} {\bibfield  {journal} {\bibinfo  {journal} {J.
  Comput. Chem.}\ }\textbf {\bibinfo {volume} {22}},\ \bibinfo {pages} {1048}
  (\bibinfo {year} {2001})}\BibitemShut {NoStop}%
\bibitem [{\citenamefont {Maier}\ \emph {et~al.}(2015)\citenamefont {Maier},
  \citenamefont {Martinez}, \citenamefont {Kasavajhala}, \citenamefont
  {Wickstrom}, \citenamefont {Hauser},\ and\ \citenamefont
  {Simmerling}}]{Maier:2015}%
  \BibitemOpen
  \bibfield  {author} {\bibinfo {author} {\bibfnamefont {J.~A.}\ \bibnamefont
  {Maier}}, \bibinfo {author} {\bibfnamefont {C.}~\bibnamefont {Martinez}},
  \bibinfo {author} {\bibfnamefont {K.}~\bibnamefont {Kasavajhala}}, \bibinfo
  {author} {\bibfnamefont {L.}~\bibnamefont {Wickstrom}}, \bibinfo {author}
  {\bibfnamefont {K.~E.}\ \bibnamefont {Hauser}}, \ and\ \bibinfo {author}
  {\bibfnamefont {C.}~\bibnamefont {Simmerling}},\ }\href {\doibase
  10.1021/acs.jctc.5b00255} {\bibfield  {journal} {\bibinfo  {journal} {J.
  Chem. Theory Comput.}\ }\textbf {\bibinfo {volume} {11}},\ \bibinfo {pages}
  {3696} (\bibinfo {year} {2015})}\BibitemShut {NoStop}%
\bibitem [{\citenamefont {Lake}\ and\ \citenamefont
  {McCullagh}(2020)}]{Lake:2020b}%
  \BibitemOpen
  \bibfield  {author} {\bibinfo {author} {\bibfnamefont {P.~T.}\ \bibnamefont
  {Lake}}\ and\ \bibinfo {author} {\bibfnamefont {M.}~\bibnamefont
  {McCullagh}},\ }in\ \href@noop {} {\emph {\bibinfo {booktitle} {Intra- and
  Intermolecular Interactions Between Non-covalently Bonded Species}}},\
  \bibinfo {editor} {edited by\ \bibinfo {editor} {\bibfnamefont {E.~R.}\
  \bibnamefont {Bernstein}}}\ (\bibinfo  {publisher} {Elsevier},\ \bibinfo
  {address} {Oxford},\ \bibinfo {year} {2020})\ Chap.~\bibinfo {chapter} {3},
  pp.\ \bibinfo {pages} {71--89}\BibitemShut {NoStop}%
\bibitem [{\citenamefont {Wu}\ \emph {et~al.}(2011)\citenamefont {Wu},
  \citenamefont {Xing}, \citenamefont {Zhou},\ and\ \citenamefont
  {Lin}}]{Wu:2011}%
  \BibitemOpen
  \bibfield  {author} {\bibinfo {author} {\bibfnamefont {W.}~\bibnamefont
  {Wu}}, \bibinfo {author} {\bibfnamefont {L.}~\bibnamefont {Xing}}, \bibinfo
  {author} {\bibfnamefont {B.}~\bibnamefont {Zhou}}, \ and\ \bibinfo {author}
  {\bibfnamefont {Z.}~\bibnamefont {Lin}},\ }\href {\doibase
  10.1186/1475-2859-10-9} {\bibfield  {journal} {\bibinfo  {journal} {Microbial
  Cell Factories}\ }\textbf {\bibinfo {volume} {10}},\ \bibinfo {pages} {9}
  (\bibinfo {year} {2011})}\BibitemShut {NoStop}%
\bibitem [{\citenamefont {Drozdov}, \citenamefont {Grossfield},\ and\
  \citenamefont {Pappu}(2004)}]{Drozdov:2004}%
  \BibitemOpen
  \bibfield  {author} {\bibinfo {author} {\bibfnamefont {A.~N.}\ \bibnamefont
  {Drozdov}}, \bibinfo {author} {\bibfnamefont {A.}~\bibnamefont {Grossfield}},
  \ and\ \bibinfo {author} {\bibfnamefont {R.~V.}\ \bibnamefont {Pappu}},\
  }\href {\doibase 10.1021/ja039051x} {\bibfield  {journal} {\bibinfo
  {journal} {J. Am. Chem. Soc.}\ }\textbf {\bibinfo {volume} {126}},\ \bibinfo
  {pages} {2574} (\bibinfo {year} {2004})}\BibitemShut {NoStop}%
\bibitem [{\citenamefont {Cai}\ \emph {et~al.}(2015)\citenamefont {Cai},
  \citenamefont {Du}, \citenamefont {Liu},\ and\ \citenamefont
  {Su}}]{Cai:2015}%
  \BibitemOpen
  \bibfield  {author} {\bibinfo {author} {\bibfnamefont {K.}~\bibnamefont
  {Cai}}, \bibinfo {author} {\bibfnamefont {F.}~\bibnamefont {Du}}, \bibinfo
  {author} {\bibfnamefont {J.}~\bibnamefont {Liu}}, \ and\ \bibinfo {author}
  {\bibfnamefont {T.}~\bibnamefont {Su}},\ }\href {\doibase
  10.1016/j.saa.2014.08.126} {\bibfield  {journal} {\bibinfo  {journal}
  {Spectrochimica Acta Part A: Molecular and Biomolecular Spectroscopy}\
  }\textbf {\bibinfo {volume} {137}},\ \bibinfo {pages} {701} (\bibinfo {year}
  {2015})}\BibitemShut {NoStop}%
\bibitem [{\citenamefont {Rubio-Martinez}, \citenamefont {Tomas},\ and\
  \citenamefont {Perez}(2017)}]{RubioMartinez:2017}%
  \BibitemOpen
  \bibfield  {author} {\bibinfo {author} {\bibfnamefont {J.}~\bibnamefont
  {Rubio-Martinez}}, \bibinfo {author} {\bibfnamefont {M.~S.}\ \bibnamefont
  {Tomas}}, \ and\ \bibinfo {author} {\bibfnamefont {J.~J.}\ \bibnamefont
  {Perez}},\ }\href {\doibase 10.1016/j.jmgm.2017.10.005} {\bibfield  {journal}
  {\bibinfo  {journal} {J.Mol. Graph. Model.}\ }\textbf {\bibinfo {volume}
  {78}},\ \bibinfo {pages} {118} (\bibinfo {year} {2017})}\BibitemShut
  {NoStop}%
\end{thebibliography}%


\begin{thebibliography}{4}%
\makeatletter
\providecommand \@ifxundefined [1]{%
 \@ifx{#1\undefined}
}%
\providecommand \@ifnum [1]{%
 \ifnum #1\expandafter \@firstoftwo
 \else \expandafter \@secondoftwo
 \fi
}%
\providecommand \@ifx [1]{%
 \ifx #1\expandafter \@firstoftwo
 \else \expandafter \@secondoftwo
 \fi
}%
\providecommand \natexlab [1]{#1}%
\providecommand \enquote  [1]{``#1''}%
\providecommand \bibnamefont  [1]{#1}%
\providecommand \bibfnamefont [1]{#1}%
\providecommand \citenamefont [1]{#1}%
\providecommand \href@noop [0]{\@secondoftwo}%
\providecommand \href [0]{\begingroup \@sanitize@url \@href}%
\providecommand \@href[1]{\@@startlink{#1}\@@href}%
\providecommand \@@href[1]{\endgroup#1\@@endlink}%
\providecommand \@sanitize@url [0]{\catcode `\\12\catcode `\$12\catcode
  `\&12\catcode `\#12\catcode `\^12\catcode `\_12\catcode `\%12\relax}%
\providecommand \@@startlink[1]{}%
\providecommand \@@endlink[0]{}%
\providecommand \url  [0]{\begingroup\@sanitize@url \@url }%
\providecommand \@url [1]{\endgroup\@href {#1}{\urlprefix }}%
\providecommand \urlprefix  [0]{URL }%
\providecommand \Eprint [0]{\href }%
\providecommand \doibase [0]{http://dx.doi.org/}%
\providecommand \selectlanguage [0]{\@gobble}%
\providecommand \bibinfo  [0]{\@secondoftwo}%
\providecommand \bibfield  [0]{\@secondoftwo}%
\providecommand \translation [1]{[#1]}%
\providecommand \BibitemOpen [0]{}%
\providecommand \bibitemStop [0]{}%
\providecommand \bibitemNoStop [0]{.\EOS\space}%
\providecommand \EOS [0]{\spacefactor3000\relax}%
\providecommand \BibitemShut  [1]{\csname bibitem#1\endcsname}%
\let\auto@bib@innerbib\@empty
\bibitem [{\citenamefont {Fox}\ and\ \citenamefont {Kollman}(1998)}]{Fox:1998}%
  \BibitemOpen
  \bibfield  {author} {\bibinfo {author} {\bibfnamefont {T.}~\bibnamefont
  {Fox}}\ and\ \bibinfo {author} {\bibfnamefont {P.~A.}\ \bibnamefont
  {Kollman}},\ }\href {\doibase 10.1021/jp9717655} {\bibfield  {journal}
  {\bibinfo  {journal} {Langmuir}\ }\textbf {\bibinfo {volume} {102}},\
  \bibinfo {pages} {8070} (\bibinfo {year} {1998})}\BibitemShut {NoStop}%
\bibitem [{\citenamefont {Cieplak}, \citenamefont {Caldwell},\ and\
  \citenamefont {Kollman}(2001)}]{Cieplak:2001}%
  \BibitemOpen
  \bibfield  {author} {\bibinfo {author} {\bibfnamefont {P.}~\bibnamefont
  {Cieplak}}, \bibinfo {author} {\bibfnamefont {J.}~\bibnamefont {Caldwell}}, \
  and\ \bibinfo {author} {\bibfnamefont {P.}~\bibnamefont {Kollman}},\ }\href
  {\doibase 10.1002/jcc.1065} {\bibfield  {journal} {\bibinfo  {journal} {J.
  Comput. Chem.}\ }\textbf {\bibinfo {volume} {22}},\ \bibinfo {pages} {1048}
  (\bibinfo {year} {2001})}\BibitemShut {NoStop}%
\bibitem [{\citenamefont {Case}\ \emph {et~al.}(2018)\citenamefont {Case},
  \citenamefont {Betz}, \citenamefont {Cerutti}, \citenamefont {{Cheatham
  III}}, \citenamefont {Darden}, \citenamefont {Duke}, \citenamefont {Giese},
  \citenamefont {Gohlke}, \citenamefont {Goetz}, \citenamefont {Homeyer},
  \citenamefont {Izadi}, \citenamefont {Janowski}, \citenamefont {Kaus},
  \citenamefont {Kovalenko}, \citenamefont {Lee}, \citenamefont {LeGrand},
  \citenamefont {Li}, \citenamefont {Lin}, \citenamefont {Luchko},
  \citenamefont {Luo}, \citenamefont {Madej}, \citenamefont {Mermelstein},
  \citenamefont {Merz}, \citenamefont {Monard}, \citenamefont {Nguyen},
  \citenamefont {Nguyen}, \citenamefont {Omelyan}, \citenamefont {Onufriev},
  \citenamefont {Roe}, \citenamefont {Roitberg}, \citenamefont {Sagui},
  \citenamefont {Simmerling}, \citenamefont {Botello-Smith}, \citenamefont
  {Swails}, \citenamefont {Walker}, \citenamefont {Wang}, \citenamefont {Wolf},
  \citenamefont {Wu}, \citenamefont {Xiao},\ and\ \citenamefont
  {Kollman}}]{Amber2018}%
  \BibitemOpen
  \bibfield  {author} {\bibinfo {author} {\bibfnamefont {D.}~\bibnamefont
  {Case}}, \bibinfo {author} {\bibfnamefont {R.}~\bibnamefont {Betz}}, \bibinfo
  {author} {\bibfnamefont {D.}~\bibnamefont {Cerutti}}, \bibinfo {author}
  {\bibfnamefont {T.}~\bibnamefont {{Cheatham III}}}, \bibinfo {author}
  {\bibfnamefont {T.}~\bibnamefont {Darden}}, \bibinfo {author} {\bibfnamefont
  {R.}~\bibnamefont {Duke}}, \bibinfo {author} {\bibfnamefont {T.}~\bibnamefont
  {Giese}}, \bibinfo {author} {\bibfnamefont {H.}~\bibnamefont {Gohlke}},
  \bibinfo {author} {\bibfnamefont {A.}~\bibnamefont {Goetz}}, \bibinfo
  {author} {\bibfnamefont {N.}~\bibnamefont {Homeyer}}, \bibinfo {author}
  {\bibfnamefont {S.}~\bibnamefont {Izadi}}, \bibinfo {author} {\bibfnamefont
  {P.}~\bibnamefont {Janowski}}, \bibinfo {author} {\bibfnamefont
  {J.}~\bibnamefont {Kaus}}, \bibinfo {author} {\bibfnamefont {A.}~\bibnamefont
  {Kovalenko}}, \bibinfo {author} {\bibfnamefont {T.}~\bibnamefont {Lee}},
  \bibinfo {author} {\bibfnamefont {S.}~\bibnamefont {LeGrand}}, \bibinfo
  {author} {\bibfnamefont {P.}~\bibnamefont {Li}}, \bibinfo {author}
  {\bibfnamefont {C.}~\bibnamefont {Lin}}, \bibinfo {author} {\bibfnamefont
  {T.}~\bibnamefont {Luchko}}, \bibinfo {author} {\bibfnamefont
  {R.}~\bibnamefont {Luo}}, \bibinfo {author} {\bibfnamefont {B.}~\bibnamefont
  {Madej}}, \bibinfo {author} {\bibfnamefont {D.}~\bibnamefont {Mermelstein}},
  \bibinfo {author} {\bibfnamefont {K.}~\bibnamefont {Merz}}, \bibinfo {author}
  {\bibfnamefont {G.}~\bibnamefont {Monard}}, \bibinfo {author} {\bibfnamefont
  {H.}~\bibnamefont {Nguyen}}, \bibinfo {author} {\bibfnamefont
  {H.}~\bibnamefont {Nguyen}}, \bibinfo {author} {\bibfnamefont
  {I.}~\bibnamefont {Omelyan}}, \bibinfo {author} {\bibfnamefont
  {A.}~\bibnamefont {Onufriev}}, \bibinfo {author} {\bibfnamefont
  {D.}~\bibnamefont {Roe}}, \bibinfo {author} {\bibfnamefont {A.}~\bibnamefont
  {Roitberg}}, \bibinfo {author} {\bibfnamefont {C.}~\bibnamefont {Sagui}},
  \bibinfo {author} {\bibfnamefont {C.}~\bibnamefont {Simmerling}}, \bibinfo
  {author} {\bibfnamefont {W.}~\bibnamefont {Botello-Smith}}, \bibinfo {author}
  {\bibfnamefont {J.}~\bibnamefont {Swails}}, \bibinfo {author} {\bibfnamefont
  {R.}~\bibnamefont {Walker}}, \bibinfo {author} {\bibfnamefont
  {J.}~\bibnamefont {Wang}}, \bibinfo {author} {\bibfnamefont {R.}~\bibnamefont
  {Wolf}}, \bibinfo {author} {\bibfnamefont {X.}~\bibnamefont {Wu}}, \bibinfo
  {author} {\bibfnamefont {L.}~\bibnamefont {Xiao}}, \ and\ \bibinfo {author}
  {\bibfnamefont {P.}~\bibnamefont {Kollman}},\ }\href@noop {} {\emph {\bibinfo
  {title} {{Amber 2018}}}}\ (\bibinfo  {publisher} {University of California,
  San Francisco.},\ \bibinfo {year} {2018})\BibitemShut {NoStop}%
\bibitem [{\citenamefont {Mongan}\ \emph {et~al.}(2007)\citenamefont {Mongan},
  \citenamefont {Simmerling}, \citenamefont {A.~McCammon}, \citenamefont
  {A.~Case},\ and\ \citenamefont {Onufriev}}]{Mongan:2007}%
  \BibitemOpen
  \bibfield  {author} {\bibinfo {author} {\bibfnamefont {J.}~\bibnamefont
  {Mongan}}, \bibinfo {author} {\bibfnamefont {C.}~\bibnamefont {Simmerling}},
  \bibinfo {author} {\bibfnamefont {J.}~\bibnamefont {A.~McCammon}}, \bibinfo
  {author} {\bibfnamefont {D.}~\bibnamefont {A.~Case}}, \ and\ \bibinfo
  {author} {\bibfnamefont {A.}~\bibnamefont {Onufriev}},\ }\href@noop {}
  {\bibfield  {journal} {\bibinfo  {journal} {. Chem. Theory Comput.}\ }\textbf
  {\bibinfo {volume} {3}},\ \bibinfo {pages} {156} (\bibinfo {year}
  {2007})}\BibitemShut {NoStop}%
\end{thebibliography}%

\end{document}



\title{Supplementary Material for: Implicit Solvation Using the Superposition Approximation (IS-SPA): Extension to Polar Solutes in Chloroform.} 



\author{Peter T. Lake}
\affiliation{Department of Chemistry, Oklahoma State University}

\author{Max A. Mattson}
\affiliation{Department of Chemistry, Colorado State University}

\author{Martin McCullagh}
\email[]{martin.mccullagh@okstate.edu}
\affiliation{Department of Chemistry, Oklahoma State University}


\date{\today}

\pacs{}

\maketitle 

\section{Simulation Methods}
\label{sim_methods}

The model parameters used for chloroform in aaMD are for a completely flexible model.\cite{Fox:1998,Cieplak:2001}  The custom-made single ion solutes were given Lennard-Jones parameters of $\epsilon=0.152$ kcal/mol and $r_{\text{min}}=7$ \AA\ and charges spanning 0.0 e to 1.0 e of equal and opposite sign.  The parameters for alanine dipeptide are from the ff14SB force field.\cite{Maier:2015}  Explicit solvent molecular dynamics simulations along with the Generalized Born (GB) and constant CDD simulations are performed with the AMBER 18 software package.\cite{Amber2018}  The GB model used is the GBneck model using the modified Bondi radii for the atoms.\cite{Mongan:2007}  The implicit solvent models use an external dielectric constant of 2.3473, as measured in the model solvent.  The IS-SPA simulations are performed using our own Fortran code.

All simulations are performed at a temperature of 298 K using a 2 fs time step in the simulation.  A Langevin thermostat with a collision frequency of 2 ps$^{-1}$ is used in the simulations using AMBER and an Andersen thermostat with a collision frequency of 16.67 ps$^{-1}$ is used for IS-SPA simulations.  The explicit solvent simulations are run at a pressure of 1 bar using a Berendsen barostat.  For the explicit solvent simulations, a direct interaction cut off of 12 \AA\ is used with particle mesh Ewald being used to account for long range electrostatic forces.  No cut offs are used in the implicit solvent simulations.  Simulations run in AMBER use SHAKE to restrain the bonds containing hydrogen atoms in the solute molecule while the IS-SPA code allows for fluctuations in these bonds but changes the mass of the hydrogen atoms to 12 u to reduce the frequency of those bonds.

Initial frames for all the implicit solvent simulations are collected from the explicit solvent simulations.  The input for the explicit solvent simulations are produced using tleap.  The energy of the initial configuration is first minimized for 20000 steps.  The system is then heated to 298 K over 50 ps and then the volume is allowed to change for another 50 ps.  The system is evolved for 3 ns before collecting data in order to remove any artifacts from the initial configuration.

The solvent forces in the IS-SPA simulations are calculated by MC integration of the integral in Equation (1) from the Theory section.  The focus of the current study is not on the efficiency of calculating this integral.  Instead, the sampling is performed in each step with 100 MC sampled points per solute atom uniformly chosen in a spherical volume around a solute atom with a 12 \AA\ radius.  The long range electrostatic force calculation discussed in the Theory section requires sampling even within the excluded volume of the solute.  For the electrostatic force, the axial multipole moments of chloroform are calculated using the minimum energy structure of the model system at a position of $d=-0.21$ \AA\ as described in the Theory section.  These values are 0.2291 e\AA\ for the dipole moment, 0.09275 e\AA$^2$ for the quadrupole moment, 0.5140 e\AA$^3$ for the octupole moment, and 0 e\AA$^4$ for the hexadecapole moment.  The calculation of the electrostatic forces is truncated at the octupole term.

The parameters for IS-SPA of the single Lennard-Jones ion are calculated from a simulation of a single ion solvated with 33,000 chloroform molecules resulting in a cubic box with an approximate linear length of 165 \AA.  A total of 10 ns of simulation is run for each system and the average solvent density, polarization, and Lennard-Jones force is measured in 0.1 \AA\ bins.

The mean forces and PMFs of the Lennard-Jones ions are calculated by performing umbrella sampling (US) simulations as a function of the separation distance between the two solutes of equal and opposite charges.  The two ions are solvated with 1,800 chloroform molecules resulting in a cubic box with an approximate linear length of 62 \AA.  A harmonic restraint with a spring constant of 20 kcal/mol/\AA$^2$ is applied to the two ions.  The direct interaction between the two ions is not calculated allowing for calculation of the PMF between 0 \AA\ and 25 \AA, using windows separated by 0.5 \AA.  Each window is simulated for 100 ns.  Only the explicit solvent model is simulated since the IS-SPA results can be obtained by direct integration of the mean force for a given ion separation.

The distribution of internal orientations of alanine dipeptide is calculated by simulating a single molecule.  For the explicit solvent simulation, the molecule is solvated with 5,000 chloroform molecules resulting in a cubic box with an approximate linear length of 65 \AA.  A bias potential is applied to the $\phi$-dihedral angle in order to fully sample the dihedral states of the molecule.  The bias potential is found by calculating the minimum energy states of the system in vacuum in all $\phi$- and $\psi$-dihedral space and finding the Fourier decomposition to third order.  The resulting potentials have spring constants of $k_1=3.8124$ kcal/mol, $k_2=3.2418$ kcal/mol, and $k_3=3.2418$ kcal/mol, and phase shifts of $\psi_1=4.6815$ rad, $\psi_2=1.9924$ rad and, $\psi_3=3.3407$ rad, where the subscript is the periodicity of the potential.  All solvent model systems were run with 5 replicas for 200 ns each.

Average solvent distributions for a single solute configuration are needed to parameterize IS-SPA.  A frame of alanine dipeptide in explicit solvent in the minimum free energy dihedral configuration is chosen for this purpose.  The solute atoms are then restrained as to not move and the solvent distribution is sampled in a simulation of 100 ns.  The resulting solvent density and polarization are then measured in cubic bins of 0.25 \AA\ length around the molecule.

The mean force and PMF for two alanine dipeptide molecules are calculated by performing US simulations as a function of separation distance from the center of mass distance between the three central heavy atoms in the solute.  In the case of the explicit solvent simulations, the two solutes are solvated in 2,000 chloroform molecules resulting in a cubic box with an approximate linear length of 65 \AA.  A harmonic restraint with a spring constant of 20 kcal/mol/\AA$^2$ is applied to the two solute molecules.  The PMF is calculated via windows between 3.5 \AA\ and 16.5 \AA\ separated by 0.5 \AA.  Each window is simulated for 100 ns.

\section{Multipole moments of chloroform}

\begin{figure}[h]
\includegraphics[width=0.8\textwidth]{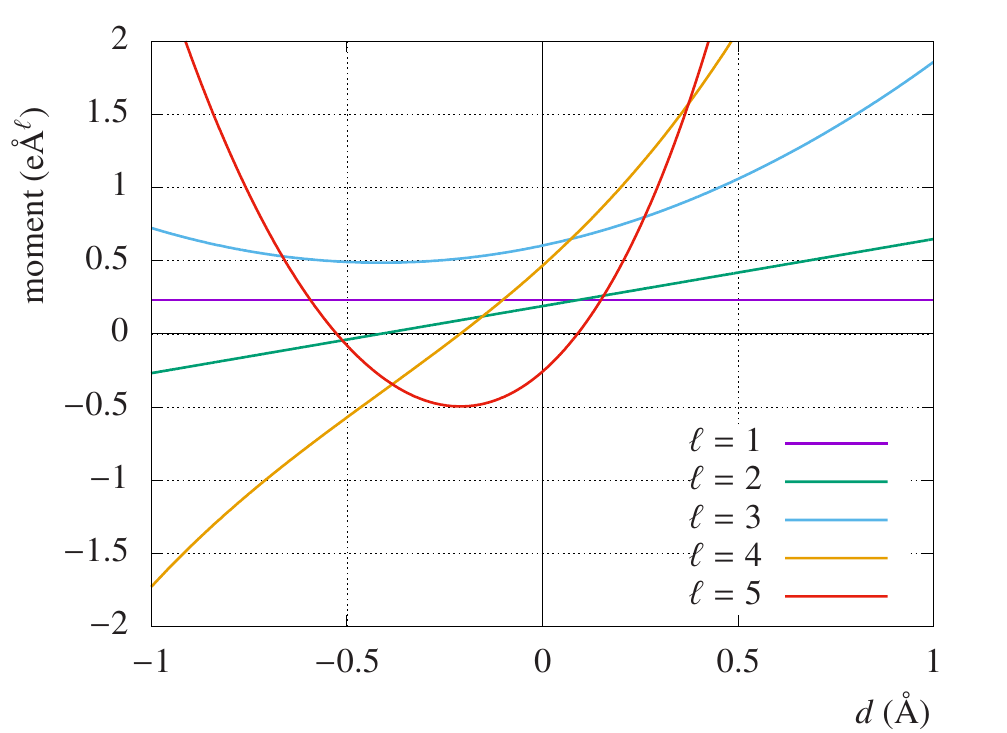}
\caption{\label{moments} Magnitude of the multipole moments for chloroform as a function of distance from the carbon atom. A positive value is towards the chlorine atoms.}
\end{figure}

\section{Electrostatic solvent forces in IS-SPA}

The force of the chloroform solvent electrostatic force is expanded in a multipole expansion.  The assumption is that the distribution of the dipole moment of a chloroform molecule at a given point in space is that of an ideal dipole.  Given that distribution, the average force is then expanded.  The force is calculated as
\begin{equation}
    \bm{f}^C_{solvent}=\sum_{\ell=1}^\infty\frac{q_iM_\ell\langle P_\ell(\bm{\hat{p}}\cdot\bm{\hat{r}})\rangle}{4\pi\epsilon_0r^{\ell+2}}\left( P'_{\ell+1}( \bm{\hat{E}}\cdot\bm{\hat{r}})\bm{\hat{r}} - P'_{\ell}(\bm{\hat{E}}\cdot\bm{\hat{r}})\bm{\hat{E}}\right),
\end{equation}
where $q_i$ is the charge on the solute atom, $M_\ell$ are the multipole moments, $\bm{\hat{r}}$ and $r$ are the unit vector and distance between the solute and solvent, $\bm{\hat{E}}$ is the unit vector of the electric field, and $P_\ell$ are the Legendre polynomials.  The angle brackets is the ensemble average over the distribution of dipole moments.  This requires calculating $\langle (\bm{\hat{r}}\cdot\bm{\hat{p}})^\ell \rangle$ for a dipole in a uniform electric field, which is calculated recursively as
\begin{equation}
    \langle (\bm{\hat{r}}\cdot\bm{\hat{p}})^\ell \rangle=-\frac{\ell T}{pE}\langle (\bm{\hat{r}}\cdot\bm{\hat{p}})^{\ell-1} \rangle + \begin{cases}
     1 & \ell \mbox{ even}\\
     \frac{1}{\tanh{\frac{pE}{T}}}& \ell \mbox{ odd}
  \end{cases},
\end{equation}
with $\langle (\bm{\hat{r}}\cdot\bm{\hat{p}})^0 \rangle=1$.

\section{Atomic parameters for alanine dipeptide}

\begin{figure}[h]
\includegraphics[width=1.\textwidth]{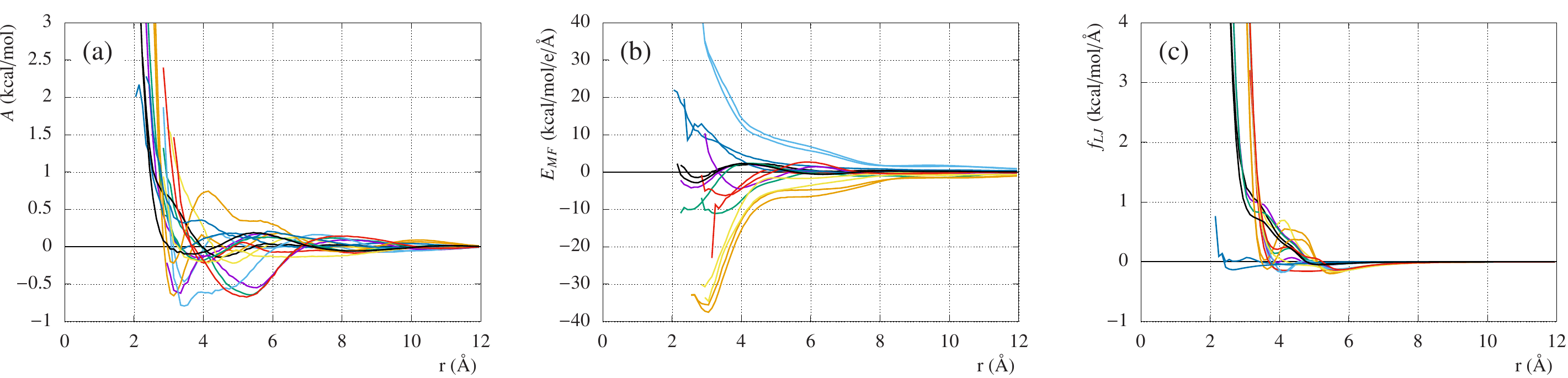}
\caption{\label{spag} Plots of the atomic parameters for each atom type in alanine dipeptide. a) The free energy of the solvent to a given atom. b) The effective electric field from a given atom.  c) The average Lennard-Jones force between a given atom and the solvent.}
\end{figure}


\bibliography{SI}